\documentclass[prl,superscriptaddress,showpacs,longbibliography,reprint]{revtex4-2} 

\usepackage[colorlinks=true,linkcolor=blue,urlcolor=blue,citecolor=blue,pdfusetitle]{hyperref}
\usepackage{color}
\usepackage[utf8]{inputenc}
\usepackage[usenames,dvipsnames]{xcolor}
\usepackage{amsthm}
\usepackage{amsmath}
\usepackage{amssymb}
\usepackage{graphicx}
\graphicspath{{graphics/}}
\usepackage{epsfig}
\usepackage{dcolumn}
\usepackage{bm}
\usepackage{mathrsfs}
\usepackage{multirow}
\usepackage[all]{xy}
\usepackage{pbox}
\usepackage{lipsum}
\usepackage{verbatim}
\usepackage{braket}
\usepackage{dsfont}
\usepackage{array}
\usepackage{makecell}
\usepackage{tabularx}
\usepackage{isotope}
\usepackage{xr}
\usepackage{float}

\newcommand{\tr}{\mathrm{tr}}

\begin{document}
\raggedbottom

\title{Designing open quantum systems for enabling quantum enhanced sensing through classical measurements}

\newcommand{\tubingen}{Institut f\"{u}r Theoretische Physik and Center for Integrated Quantum Science and Technology,  Universit\"{a}t T\"{u}bingen, Auf der Morgenstelle 14, 72076 T\"{u}bingen, Germany}

\author{Robert Mattes}
\affiliation{\tubingen}

\author{Albert Cabot}
\affiliation{\tubingen}

\author{Federico Carollo}
\affiliation{Centre for Fluid and Complex Systems, Coventry University, Coventry, CV1 2TT, United Kingdom}

\author{Igor Lesanovsky}
\affiliation{\tubingen}
\affiliation{School of Physics and Astronomy and Centre for the Mathematics and Theoretical Physics of Quantum Non-Equilibrium Systems, The University of Nottingham, Nottingham, NG7 2RD, United Kingdom}

\date{\today}

\begin{abstract}

Quantum systems in nonequilibrium conditions, where coherent many-body  interactions compete with dissipative effects, can  feature rich phase diagrams and emergent critical behavior. Associated collective effects, together with the continuous observation of quanta dissipated into the environment --- typically photons --- allow to achieve quantum enhanced parameter estimation. However, protocols for tapping this enhancement typically involve intricate measurements on the combined system-environment state. Here we show that  many-body quantum enhancement can in fact be obtained through  classical measurements, such as photon counting and homodyne detection. We illustrate this in detail for a class of open spin-boson models which can be realized in trapped-ion or cavity QED setups. Our findings  highlight a route towards the design of systems that enable a practical implementation of quantum enhanced metrology through continuous classical measurements.

\end{abstract}

\maketitle

The interplay between collective interactions and dissipation in open quantum systems gives rise to rich phase diagrams hosting genuine nonequilibrium many-body phenomena. These include superradiant~\cite{dicke_coherence_1954,hepp_equilibrium_1973,gross_superradiance_1982,bastidas_nonequilibrium_2012,Bohnet2012,Norcia2016,Ferioli2023} or time-crystal phases~\cite{wilczek_quantum_2012,sacha_time_2017,iemini_boundary_2018,Gong2018,Gambetta2019,Zhu2019,booker_non-stationarity_2020,kongkhambut_observation_2022,Cabot2023,Mattes2023,Solanki2024}. The associated collective behavior is typically characterized by divergent correlations, which may be useful for metrological applications~\cite{Fernandez2017,Garbe2020,montenegro_quantum_2023,montenegro_review_2024,gietka_understanding_2022,ding_enhanced_2022,ilias_criticality-enhanced_2022,Alushi2025,midha_metrology_2025,lee_timescales_2025}. In these  settings, part of the information about a system parameter to be determined is emitted into the environment. The ultimate sensitivity is therefore only achieved, in principle, by considering the joint system-environment state~\cite{gammelmark_fisher_2014,macieszczak_dynamical_2016,albarelli_restoring_2018,khan_tensor_2025,yang_quantum_2025}. This sensitivity poses bounds, for instance, to the accuracy that is achievable by exploiting  protocols based on continuous monitoring of the output of the system alone~\cite{Catana2012,gammelmark_bayesian_2013,kiilerich_estimation_2014,kiilerich_bayesian_2016,Martinez_kalman_2018,Shankar2019,rossi_noisy_2020,radaelli_parameter_2024,Rinaldi2024,Cabot2024_symmetries}.  The latter are  generally not capable of extracting the full information and one typically needs to resort to advanced measurement schemes, that might be challenging to implement in practice~\cite{yang_efficient_2023,zhou_saturating_2020,godley_adaptive_2023,cabot_continuous_2024}.
\begin{figure}[t!]
    \centering
    \includegraphics[width=\columnwidth]{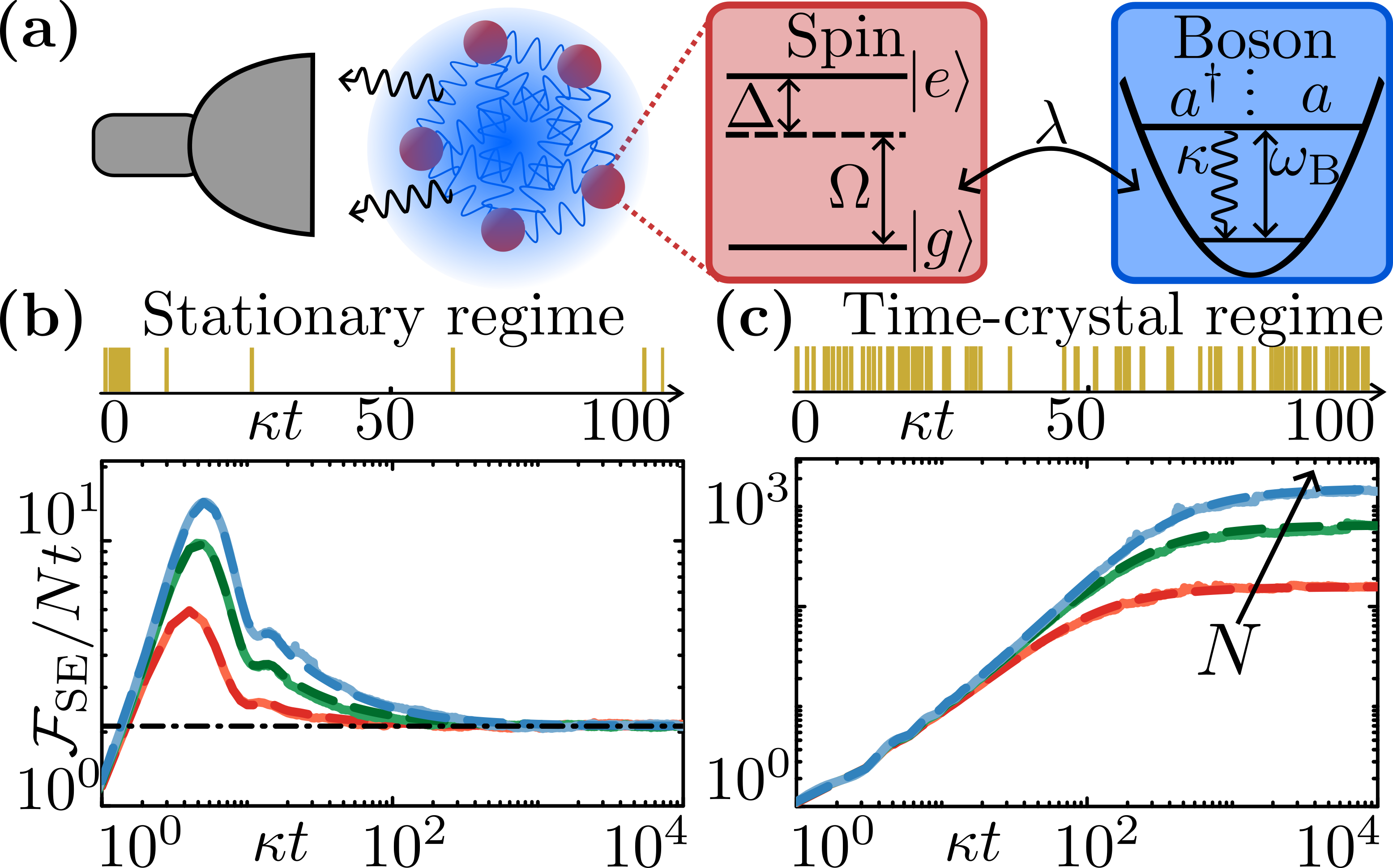}
    \caption{{\bf Parameter estimation with open many-body system via continuous monitoring.} (a) To illustrate our ideas we consider spin-boson models, which consist of $N$ two-level atoms, with ground state $\ket{g}$ and excited state $\ket{e}$, driven by a laser (Rabi frequency $\Omega$ and detuning $\Delta$) and collectively interacting with a bosonic mode (detuned from the laser by $\delta=\omega_{\mathrm{B}} - \omega_{\mathrm{L}}$). Decay of bosonic excitations into the environment (rate $\kappa$) are continuously monitored. 
    Typical photocounting emission record in the stationary (b) $\Omega/\kappa =0.1$ and in the time-crystal regime (c) $\Omega/\kappa=2$ of the Tavis-Cummings model [cf. Eq.~\eqref{eq:TC_Hamiltonian}], with $\lambda/\kappa=0.5$.
    QFI of system-environment state (dashed) and QFI for photocounting trajectories (solid), for sensing the Rabi frequency $\Omega$, averaged over $10^4$ trajectories, for system sizes of $N=5,10,15$. The initial state features all spins in the excited state and the bosonic mode in the vacuum. In panel (b) the dashed-dotted line represents the analytical value of the system-environment QFI in the stationary regime~\cite{SM}.}
    \label{Fig1}
\end{figure}

In this work, we show that classical measurements, such as photon counting and homodyne detection, can in fact be optimal. This means that under certain conditions - which we derive explicitly - they allow to tap many-body quantum enhancement. We demonstrate this exemplarily for spin-boson systems, cf.~Fig.~\ref{Fig1}, which are minimal models for a wide range of quantum optical  systems~\cite{mivehvar_cavity_2021,norcia_cavity-mediated_2018,dogra_dissipation-induced_2019,Muniz2020,porras_mesoscopic_2008,barreiro_open-system_2011,blatt_quantum_2012,Shankar2017,lemmer_trapped-ion_2018,Gilmore2021,mei_experimental_2022,sun_quantum_2025}. Our results explain and generalize findings of previous studies on single- or few-body systems~\cite{gammelmark_bayesian_2013,kiilerich_estimation_2014,kiilerich_bayesian_2016,radaelli_parameter_2024} and approaches relying on the construction of dark stationary states~\cite{lee_timescales_2025}, in which homodyne detection or photocounting monitoring were found to be optimal. The many-body models we consider show collective quantum-enhanced sensitivity witnessed by the scaling, with the system size, of the \textit{Quantum Fisher information} (QFI)~\cite{braunstein_statistical_1994} of the combined system-environment state. While previous studies on similar models rely on protocols for extracting this many-body enhancement involving a replica system acting as a coherent absorber~\cite{cabot_continuous_2024}, here we demonstrate that the \textit{classical Fisher information} (FI) of the photocount record or of the homodyne current~\cite{gammelmark_bayesian_2013,albarelli_restoring_2018,radaelli_parameter_2024,khan_tensor_2025} are sufficient to recover the QFI of the compound system-environment state in the long-time limit. A possible advantage of such approach is that information about a parameter is acquired gradually and continuously with time. This contrasts other protocols relying on the preparation of correlated quantum states and strong (projective) measurements~\cite{yu_quantum_2022}. 

Our analytical results illustrate the fundamental mechanism behind the saturation of the system-environment QFI through the classical Fisher information associated with these measurement schemes of the emission field.
This unveils general principles for designing optimal open quantum sensors as well as to discern which coherent or incoherent processes are detrimental for such an optimality. Our work therefore bridges the gap between the theoretical ultimate sensitivity given by the system-environment QFI and the one experimentally accessible with classical measurements of the emissions.\\

\noindent \textbf{Open quantum dynamics.---}
We consider many-body quantum systems whose reduced-state dynamics is governed by the master equation
\begin{equation}\label{eq:master_equation}
    \dot{\rho} = -i[H,\rho] + \mathcal{D}[\rho]\, ,
\end{equation}
with $\rho$ being the reduced system state, $H$ the system Hamiltonian, and
\begin{equation}\label{eq:Dissipator}
    \mathcal{D}[\bullet] = \left(L\bullet L^\dagger - \frac{1}{2}\{L^\dagger L, \bullet\}\right)\, ,
\end{equation}
modelling dissipative effects emerging from the interaction with the environment~\cite{gorini1976,lindblad1976,breuer2002theory}. By exploiting an input-output formalism, which holds within the same approximations for which   Eq.~\eqref{eq:master_equation} is valid, it is possible to account for the evolution of  the combined system-environment state. In this approach, one considers infinitesimally small (discrete) time steps $\Delta t$ and assumes that the system interacts, at each time step, with an independent instance of the environment~\cite{gammelmark_fisher_2014,yang_efficient_2023,cabot2025}. Using this framework, one finds that the   system-environment  state is, up to a time $t=M\Delta t$, 
\begin{equation}\label{eq:pure_sys_env}
    \ket{\Psi(t)} = {\sum}_{{\bf i_t}}  K_{i_M} \dots K_{i_1} \ket{\psi(0)} \otimes \ket{i_M, \dots, i_1}\, .
\end{equation}
This expression accounts for the superposition of all possible quantum trajectories, with the system being initially in a pure state, $|\psi(0)\rangle$, and the input field in the vacuum state.
A single quantum trajectory of length $t=M\Delta t$ is described by the record ${\bf{i_t}}=\{i_1,\dots,i_M\}$ containing the states of the environmental instances at each time step after the interaction with the system. For photon counting, we have $i_j=1$ if a photon is detected at time $j\Delta t$ and $i_j=0$ otherwise \cite{Wiseman2009}. The Kraus operators $K_{i_j}$ are for the photocounting unravelling defined as
\begin{equation}\label{eq:Kraus_ops}
    K_{0} = e^{-i\left(H-\frac{i}{2}L^\dagger L\right)\Delta t}\, ,\quad  K_{1} = \sqrt{\Delta t} L\, ,
\end{equation}
where we have assumed a single decay channel with associated jump operator $L$~\footnote{This is not a fundamental limitation and a pure system-environment joint state can be similarly constructed in the presence of multiple decay channels.}. Each of the terms in Eq.~\eqref{eq:pure_sys_env} corresponds to the tensor product of the (unnormalized) system conditional state $|\tilde{\psi}_{\bf{i_t}}\rangle=K_{i_M}\dots K_{i_1}|\psi(0)\rangle$ with the corresponding measured outcome $|i_M,\dots ,i_1\rangle$ represented in a discrete time-bin mode basis~\cite{gammelmark_fisher_2014,macieszczak_dynamical_2016,yang_efficient_2023}.  The probability of each trajectory is given by the square  norm of  $|\tilde{\psi}_{\bf{i_t}}\rangle$, which allows one to obtain the corresponding normalized state $|\psi_{\bf{i_t}}\rangle$ \cite{Wiseman2009}.
Analogously, Eq.~\eqref{eq:pure_sys_env} can also be  written as the superposition of all possible homodyne trajectories (see Supplemental Material (SM)~\cite{SM}), yielding the Kraus operators
\begin{equation}\label{eq:Kraus_homodyne}
K_{J_j}=\mathbf{1}-iH\Delta t-\frac{L^\dagger L}{2}\Delta t+ L e^{i\Phi} J_j \Delta t,    
\end{equation}
where $J_j$ is the homodyne current at time $t_j=j\Delta t$. This is a stochastic quantity whose average follows the (conditional) expectation value of the monitored quadrature $(Le^{i\Phi}+L^\dagger e^{-i\Phi})$ \cite{Wiseman2009}. A single homodyne trajectory is characterized by the measured homodyne current, $J_j\in \mathbb{R}$, at each time step. We also denote the homodyne measured record and conditional state with ${\bf i_t}=\{J_1,\dots,J_M\}$ and $|\psi_{\bf i_t}\rangle$, respectively. Notice that for both unravellings by performing the continuous time limit $\Delta t\to 0$ and tracing out the environment degrees of freedom, one recovers the master equation~\eqref{eq:master_equation} for the reduced system state $\rho=\tr_{\rm{env}}\left(\ket{\Psi}\!\bra{\Psi}\right)$~\cite{gammelmark_fisher_2014,Wiseman2009}.\\

\noindent \textbf{Optimal sensing with classical measurements.---}
We are interested in estimating a dynamical parameter of the open quantum system introduced above. Such a parameter, generally denote by $\eta$, is naturally imprinted in the reduced system state $\rho$ as well as in the joint system-environment state $\ket{\Psi}$, during the evolution encoded in Eq.~\eqref{eq:master_equation} and Eq.~\eqref{eq:pure_sys_env}, respectively. However, only the  state $\ket{\Psi}$ contains the ultimate information about the parameter and its corresponding QFI is given by~\cite{gammelmark_fisher_2014}
\begin{equation}\label{eq:ultimate_Fisher}
    \mathcal{F}_{\mathrm{SE}}(\eta,t) = 4( \braket{\partial_\eta \Psi(t)|\partial_\eta \Psi(t)} + (\braket{\partial_\eta \Psi(t)|\Psi(t)})^2)\, ,
\end{equation}
where $\partial_\eta \Psi(t) = \partial\Psi(t)/(\partial \eta )$ captures the change and thus distinguishability from $\Psi(t)$ for variations of the parameter $\eta$. The variance of any (unbiased) estimator $\hat{\eta}$ of the parameter $\eta$ is lower bounded by the quantum Cramér-Rao bound (QCRB) $\mathrm{Var}(\hat{\eta}) \geq [\mathcal{F}_{\mathrm{SE}}(\eta,t)]^{-1}$~\cite{braunstein_statistical_1994,paris_quantum_2011}. While the state of the environment is generally a complicated object, the system-environment QFI given in Eq.~\eqref{eq:ultimate_Fisher} can be, in principle, computed through a two-sided master equation that involves only the system degrees of freedom~\cite{gammelmark_fisher_2014,macieszczak_dynamical_2016,SM}.
Notice that in the presence of unmonitored decay channels, e.g., due to non-ideal detection, the QFI of the monitored channels can nevertheless be obtained (see Refs.~\cite{khan_tensor_2025,yang_quantum_2025}), and is generally smaller than the one in Eq.~\eqref{eq:ultimate_Fisher}.

The QFI in Eq.~\eqref{eq:ultimate_Fisher} provides the ultimate bound in the estimation of the parameter $\eta$ using the joint system-environment state. From the structure of the state $\ket{\Psi(t)}$ [cf.~Eq.~\eqref{eq:pure_sys_env}], the optimal measurement, i.e., the one able to saturate the bound, might involve measuring joint system-environment operators. This is, in general, not  feasible experimentally and thus a fundamental challenge is to understand how the potential of the QFI can be tapped. In this work, we approach this issue from a different perspective: we fix the measurements to be either photon counting or homdyne detection, see Fig.~\ref{Fig1}(a), and ask for which systems this is optimal. As we show in the following, there exists a class of models and states for which these measurement schemes are optimal and their Fisher information saturates the QFI given in Eq.~\eqref{eq:ultimate_Fisher}. 

Continuously monitoring the system until time $t$ allows one to estimate the parameter $\eta$ by considering the measured record (photon counts or homodyne current) and taking a final direct measurement on the conditional system state~\cite{albarelli_restoring_2018}. The achievable precision with this protocol is bounded by the inverse of the following Fisher information~\cite{albarelli_restoring_2018}
\begin{equation}\label{eq:Fisher_conditional}
    \mathcal{F}_{\mathrm{SE}}^{(u)} (\eta,t)= \mathcal{I}_{\mathrm{E}}^{(u)}(\eta,t) + \mathcal{F}_{\mathrm{S}}^{(u)}(\eta,t)\, ,
\end{equation}
where $u=c,h$ denotes the two possible unravellings: photon counting or homodyne detection, respectively. $\mathcal{I}_{\mathrm{E}}^{(u)}(\eta,t)$ is the classical Fisher information associated with the probability distribution of the measured record up to time $t$, and $\mathcal{F}_{\mathrm{S}}^{(u)}$ is the average of the QFI for the conditional state for each trajectory up to time $t$. For long observation times, the contribution of the measured record  $\mathcal{I}_{\mathrm{E}}^{(u)}$ becomes proportional to $t$. $\mathcal{F}_{\mathrm{SE}}^{(u)}$ is therefore in the long-time limit dominated by $\mathcal{I}_{\mathrm{E}}^{(u)}$, and saturation (per unit of time) of  $\mathcal{F}_{\mathrm{SE}}$ by  $\mathcal{F}_{\mathrm{SE}}^{(u)}$ can be achieved without resorting to a final measurement on the system conditional state.
In particular, we find that the system-environment QFI is saturated by the one of the trajectories, i.e. $\mathcal{F}_{\mathrm{SE}}^{(u)} (\eta,t)= \mathcal{F}_{\mathrm{SE}}(\eta,t)$, when the following conditions are satisfied:
\begin{equation}\label{eq:conditions_saturation}
\begin{split}
    \text{(I)}\,\, &\text{Tr}\big[(\partial_\eta H) \rho(s)\big] = 0\quad \forall s\leq t,\\
    \text{(II)}\,\,&\braket{{\psi}_{\bf{i_t}}|\phi_{\bf{i_t}}}\in\mathbb{R}\quad \forall \,{\bf i_t},
\end{split}    
\end{equation}
where $\ket{{\phi}_{\bf{i_t}}}=\ket{\partial_\eta \tilde{\psi}_{\bf{i_t}}}/\sqrt{\braket{\tilde{\psi}_{\bf{i_t}}|\tilde{\psi}_{\bf{i_t}}}}$ and $\partial_\eta H = \partial H/(\partial \eta )$ (see SM~\cite{SM}). These conditions apply to the photocounting and homodyne unravellings defined in Eqs.~\eqref{eq:Kraus_ops} and~\eqref{eq:Kraus_homodyne}. Interestingly, the conditions in Eq.~\eqref{eq:conditions_saturation} are fundamentally related to results derived in Ref.~\cite{braunstein_statistical_1994} for generic measurements on pure states, see SM~\cite{SM} for details. In the following, we consider for the sake of concreteness two different spin-boson models, and we show that the conditions for saturation given by Eq.~\eqref{eq:conditions_saturation} are met by a wide class of states, which are accessed during the monitoring process.\\ 

\noindent \textbf{Application to spin-boson models.---}
In order to benchmark our previous derivations we consider $N$ (two-level) spin-particles collectively interacting with a bosonic mode [cf. Fig.~\ref{Fig1}], which gives rise to a variety of intriguing physical phenomena. Such scenario naturally emerges in cavity-QED settings~\cite{mivehvar_cavity_2021,norcia_cavity-mediated_2018,dogra_dissipation-induced_2019,Muniz2020}, but can also be engineered in other platforms, such as trapped-ion experiments~\cite{porras_mesoscopic_2008,barreiro_open-system_2011,blatt_quantum_2012,Shankar2017,lemmer_trapped-ion_2018,Gilmore2021,mei_experimental_2022,sun_quantum_2025}. The transition between the ground state $\ket{g}$ and the excited state $\ket{e}$ is driven by a laser with Rabi frequency $\Omega$ and detuning $\Delta$. The bosonic mode, with creation and annihilation operators $a^\dagger$ and $a$, respectively, is detuned from the laser by $\delta$. The Hamiltonian of the system, in the frame rotating with the laser frequency,  reads
\begin{equation}\label{eq:Hamiltonian}
    H = \frac{\Omega}{2} (S_+ + S_-) + \Delta S_z + \delta a^\dagger a + H_{\mathrm{SB}}\, , 
\end{equation}
where the term $H_{\mathrm{SB}}$ accounts for the collective interactions between the spin ensemble and the bosonic mode. The spin operators are defined as $S_{c} = \frac{1}{2}\sum_{k=1}^N \sigma_{c}^{(k)}$ ($c = x,y,z$) with $\sigma_{c}^{(k)}$ being the Pauli matrices of atom $k$ and $S_\pm = S_x \pm i S_y$. The bosonic operators $a$ and $a^\dagger$ satisfy the canonical commutation relation $[a,a^\dagger] = 1$. Spontaneous decay of bosonic excitations is described through the Lindblad dissipator~\eqref{eq:Dissipator} with $L=\sqrt{\kappa} a$ and $\kappa$ being the decay rate.

We consider two types of spin-boson coupling. The first one corresponds to the Tavis-Cummings coupling~\cite{tavis_exact_1967,Tavis1969,kirton_introduction_2019}
\begin{equation}\label{eq:TC_Hamiltonian}
    H_{\mathrm{SB}} =  \frac{\lambda}{\sqrt{S}}(aS_++a^\dagger S_-)\, .
\end{equation}
This interaction is rescaled by the total angular momentum $S = N/2$ in order to guarantee a well-defined thermodynamic limit for the system~\cite{SM}. For $\Delta=\delta=0$, this model features a phase transition from a stationary to a time-crystal phase~\cite{Mattes2023}. The latter nonequilibrium phase is characterized by oscillations, whose lifetime increases with system size and eventually becomes infinite in the thermodynamic limit~\cite{iemini_boundary_2018}, where a mean-field ansatz becomes  exact~\cite{kirton_introduction_2019,carollo_exactness_2021}. The second type of coupling is the generalized Dicke one~\cite{genway_generalized_2014,Kirton_suppressing_2017,kirton_introduction_2019}
\begin{equation}\label{eq:GD_Hamiltonian}
    H_{\mathrm{SB}} =  \frac{\lambda}{\sqrt{S}}(a+a^\dagger)S_z\, .
\end{equation}
For $\Delta=0$ and $\delta>0$ the system displays a dissipative superradiant phase transition in the thermodynamic limit. At the critical coupling $\lambda_c = \sqrt{[(\delta^2+(\kappa/2)^2)\Omega]/(2\delta)}$, the system passes from a phase in which the bosonic mode features few bosonic excitations to one  characterized by its  macroscopic occupation ~\cite{genway_generalized_2014,boneberg_quantum_2022} (see SM~\cite{SM}).

\begin{figure}[t]
    \centering
    \includegraphics[width=\columnwidth]{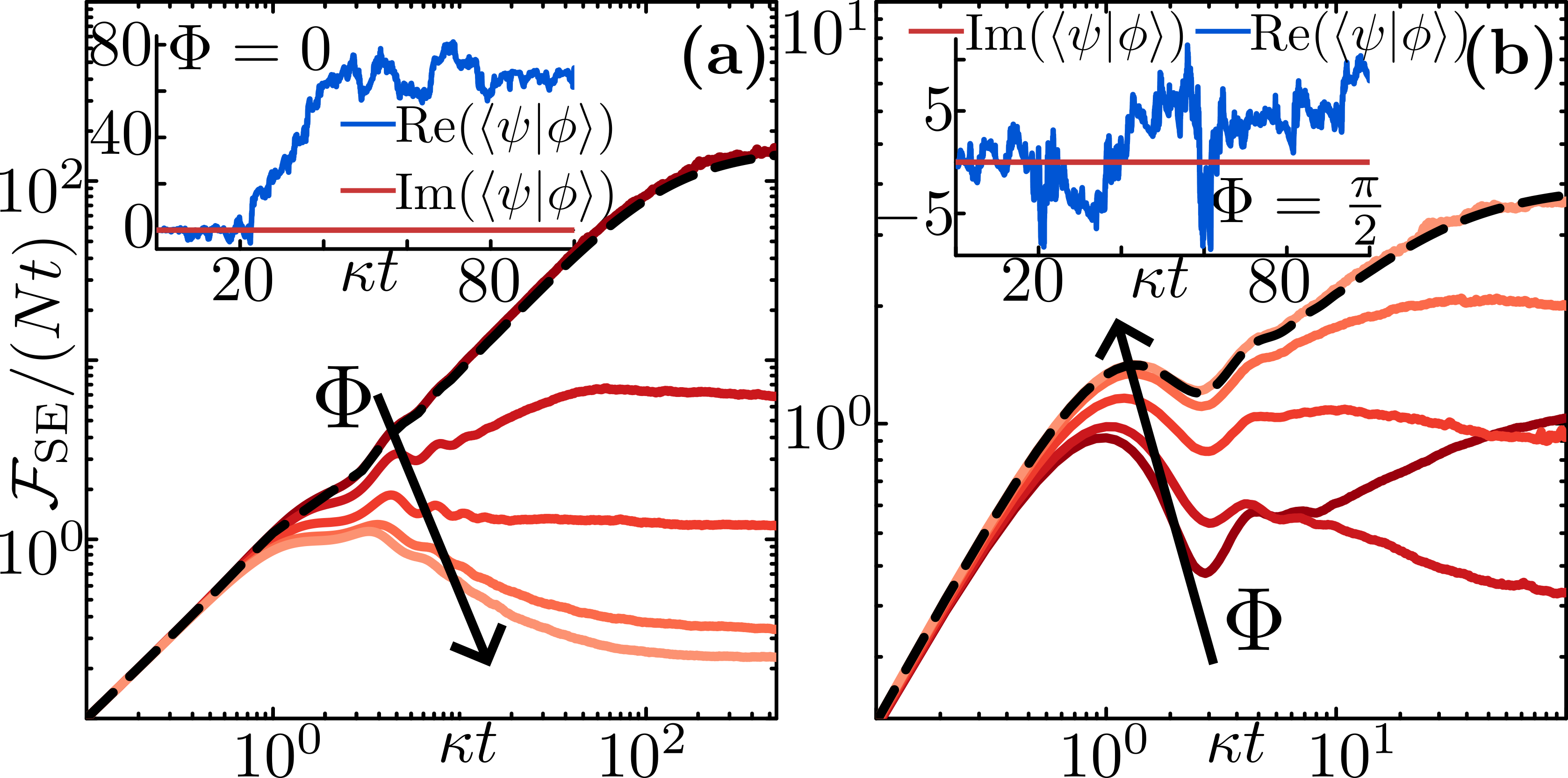}
    \caption{{\bf Saturation with homodyne measurements.} System-environment QFI (dashed) and QFI for the homodyne unravelling (solid) averaged over $2\times10^4$ trajectories with $\Phi=0,\pi/8,\pi/4,3\pi/8,\pi/2$. (a) Sensing $\Omega$ in the time-crystal regime of the Tavis-Cummings model with $N=5$, $\Omega/\kappa=2$ and $\lambda/\kappa=0.5$.
    (b) Sensing $\lambda$ in the critical regime of the generalized Dicke model with $N=5$, $\Omega/\kappa=1$, $\lambda/\kappa=1$ and $\delta=0$.
    The initial state is $\ket{\psi(0)} = \ket{S,S}\otimes\ket{0}$. The saturation condition [cf. Eq.~\eqref{eq:conditions_saturation}] is illustrated in the insets for a trajectory with $N=5$.}
    \label{Fig2}
\end{figure}
The collective phenomena emerging in these models can be used within the framework of continuous sensing.
The Tavis-Cummings coupling~\eqref{eq:TC_Hamiltonian} features a many-body enhanced sensitivity in the time-crystal regime when sensing the Rabi frequency $\Omega$, which is clearly illustrated in Fig.~\ref{Fig1}(c). This is in contrast to the stationary regime of the model [cf.~Fig.~\ref{Fig1}(b)] where, after a transient  regime, the QFI scales  linearly in time and system size~\cite{SM}.
Concerning the generalized Dicke coupling~\eqref{eq:GD_Hamiltonian}, we observe many-body enhancement only in the vicinity of the critical point, both when sensing  $\lambda$ and $\Omega$~\cite{SM}. Apart from the critical regime, it seems that a time-crystal phase is more useful for sensing the Rabi frequency $\Omega$ than the superradiant phase.

For the Tavis-Cummings coupling~\eqref{eq:TC_Hamiltonian} we now illustrate what the abstract conditions in Eq.~\eqref{eq:conditions_saturation} explicitly mean and how saturation of the system-environment QFI appears. To this end, we use Dicke~\cite{carmichael_analytical_1980} and number states as a basis for the system conditional state:
$$
   \ket{\psi} = \sum_{M_z = -S}^S \sum_{n=0}^\infty  C_{M_z,n} (\ket{S, M_z}\otimes\ket{n})\,.
$$
Let us first consider a subset of states such that they have a nonzero span over a set of Dicke and number states $\{M_z,n\}$ with consecutive z-magnetization (boson number) quantum numbers. A state of this type belongs to the class of {\it saturating states} for this spin-boson model when its coefficients satisfy:
\begin{equation}\label{eq:condition_coefficents_TC}
    C_{M_z,n}^* \cdot C_{M_z+1,n} \in \mathbb{I}\, \quad \mathrm{and} \quad C_{M_z,n}^* \cdot C_{M_z,n+1} \in \mathbb{R}\,.
\end{equation}
Intuitively, condition (I) is satisfied because the expectation values of $S_-$ and $a^\dagger S_-$ are purely imaginary for saturating states, which makes the expectation value of $\partial_{\eta}H$ with $\eta=\Omega,\lambda$ zero for all times. Instead, the fulfillment of condition (II) can be  understood from the expression of $|\partial_\eta \tilde{\psi}_{{\bf i_t}}\rangle$ in terms of the Kraus operators~\cite{SM}.
The photocounting and homodyne quantum trajectories for this model satisfy two crucial properties with regards to this class of states~\cite{SM}: (\textit{i})  saturating states evolve into saturating states; (\textit{ii}) states involving just one $|S,M_z\rangle\otimes|n\rangle$ evolve into saturating states after a single application of $K_0$ or $K_J$. For the homodyne case, these are fulfilled only when fixing $\Phi=0$, as can be understood by applying $K_J$ to a state of the type of Eq.~\eqref{eq:condition_coefficents_TC} [cf.~Fig.~\ref{Fig2}(a)].
As we show in the SM~\cite{SM} these conditions ensure that, if the initial state $|\psi(0)\rangle$ belongs to this class or is of the form $|S,M_z\rangle\otimes|n\rangle$, the saturation conditions given in Eq.~\eqref{eq:conditions_saturation} are satisfied at all times, and hence $\mathcal{F}_{\mathrm{SE}}^{(u)}(\eta,t)=\mathcal{F}_{\mathrm{SE}}(\eta, t)$, $\forall t$ and $\eta=\Omega,\lambda$. We have also numerically observed that the quantum trajectories associated with the Tavis-Cummings coupling~\eqref{eq:TC_Hamiltonian} satisfy a third relevant property (\textit{iii}): any initial condition eventually evolves into a saturating state through the action of the photocounting or homodyne measurement process. In such cases, $\mathcal{F}_{\mathrm{SE}}^{(u)}$ does not saturate the full $\mathcal{F}_{\mathrm{SE}}$. However, for long-times, saturation of the system-environment QFI rate is still observed, i.e. $\lim_{t\to\infty}[\mathcal{F}_{\mathrm{SE}}^{(u)}(\eta,t)-\mathcal{F}_{\mathrm{SE}}(\eta, t)]/t=0$ for $\eta=\Omega,\lambda$~\cite{SM}. 

Given that $\mathcal{F}_{\mathrm{SE}}^{(u)}$ is dominated by $\mathcal{I}_{\mathrm{E}}^{(u)}$ for sufficiently long times, we conclude that for the Tavis-Cummings coupling~\eqref{eq:TC_Hamiltonian} photon counting and homodyne detection are able to saturate the ultimate sensitivity rate for estimating $\Omega$ and $\lambda$. We remark that this is not a many-body effect but rather an effect rooted in the structure of the associated Kraus operators. Thus, saturation appears for any system size and the many-body enhancement of the QFI associated with the Rabi frequency $\Omega$ in the time-crystal regime is accessible with both measurement schemes. In contrast, these are no longer optimal when considering non-zero detunings, $\delta\neq0$ or $\Delta \neq 0$. 
This can be readily verified by  the application of a single Kraus operator to a saturating state, which brings it outside the class of states and thus prevents the fulfillment of conditions Eq.~\eqref{eq:condition_coefficents_TC}.
This assessment can be performed for other coherent and incoherent processes, which provides a guiding principle for designing spin-boson models for which photon counting or homodyne measurements are optimal.

\begin{figure}[t]
    \centering
    \includegraphics[width=\columnwidth]{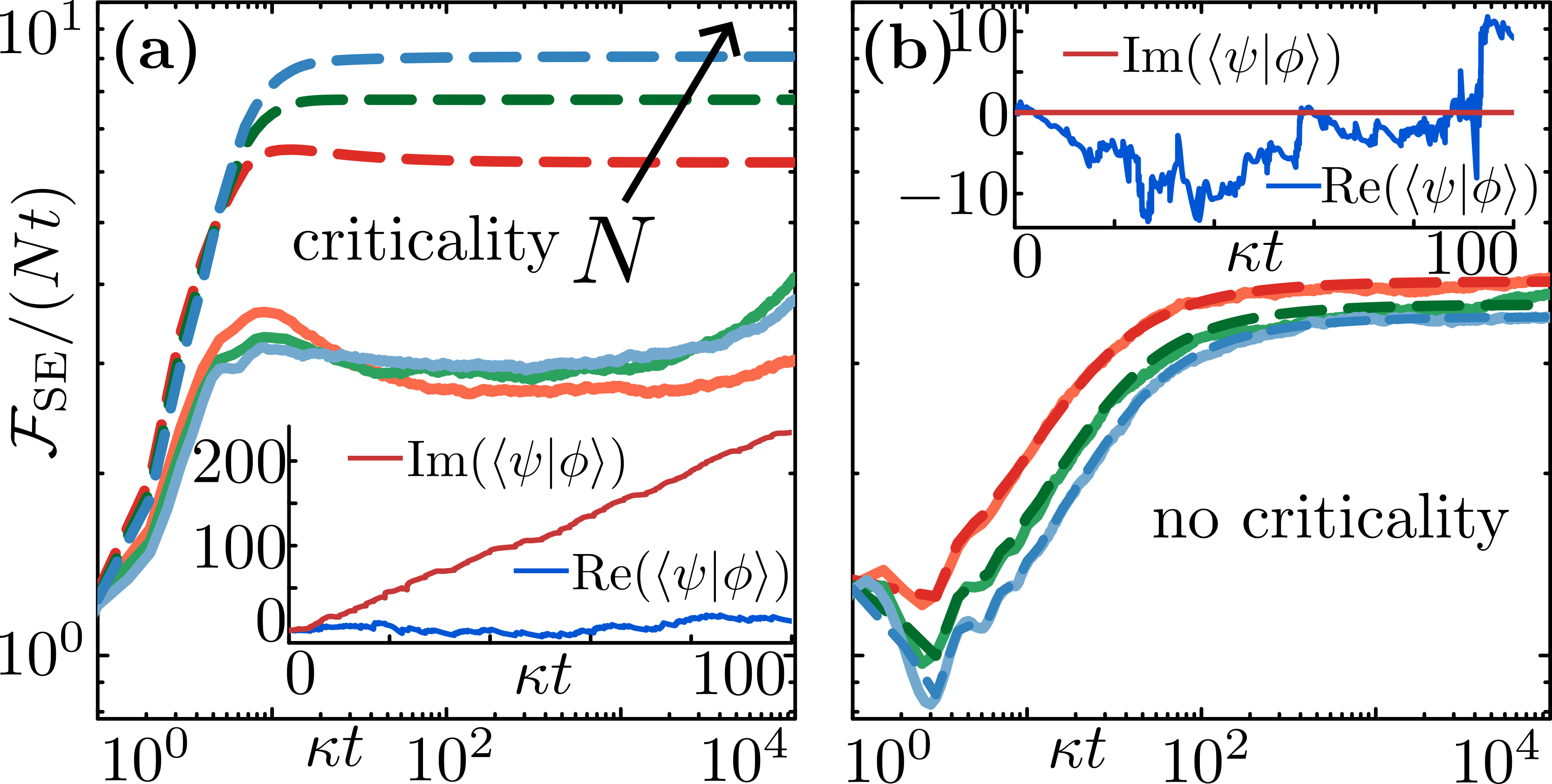}
    \caption{{\bf Saturation in the generalized Dicke model with photon counting.} System-environment QFI (dashed) and QFI for the photocounting unravelling (solid) averaged over $10^4$ trajectories for (a) $\delta/\kappa=1$  and (b) $\delta=0$, with $\Omega/\kappa=1$,  $\lambda/\kappa=1$, $N=5,7,9$ and sensing $\lambda$. The initial state is $\ket{\psi(0)} = \ket{S,S}\otimes\ket{0}$. The condition for saturation [cf. Eq.~\eqref{eq:conditions_saturation}] is illustrated in the insets for a trajectory with $N=5$.}
    \label{Fig3}
\end{figure}
For the generalized Dicke coupling~\eqref{eq:GD_Hamiltonian} we can proceed analogously, finding that the conditions in  Eq.~\eqref{eq:condition_coefficents_TC} must be replaced by
\begin{equation}\label{eq:condition_coefficents_GD}
    C_{M_z,n}^* \cdot C_{M_z+1,n} \in \mathbb{I}\, \quad \mathrm{and} \quad C_{M_z,n}^* \cdot C_{M_z,n+1} \in \mathbb{I}\, .
\end{equation}
The photocounting and homodyne (with $\Phi=\pi/2$) trajectories for this model also satisfy properties (\textit{i}) and (\textit{ii}), again only when $\delta = 0$. Conditions (I)-(II) in Eq.~\eqref{eq:conditions_saturation} are thus satisfied if the system is initially prepared in a saturating state or in a single Dicke-number state (see SM~\cite{SM}), leading to $\mathcal{F}_{\mathrm{SE}}^{(u)}(\eta,t)=\mathcal{F}_{\mathrm{SE}}(\eta, t)$ for $\eta=\Omega,\lambda$ for all times [cf. Fig.~\ref{Fig3}].
The detrimental effect of the detuning and $\Phi\neq \pi/2$, see Fig.~\ref{Fig2}(b), on optimality can be again understood from the application of the Kraus operators~\cite{SM}.  For the generalized Dicke coupling~\eqref{eq:GD_Hamiltonian}, property (\textit{iii}) is only fulfilled by homodyne measurements with $\Phi=\pi/2$. Hence saturation of the long-time QFI rate can, for this measurement, be achieved for all initial conditions~\cite{SM}.
For this model, many-body enhancement of the QFI is only found at the critical value of the coupling strength [cf.~Eq.~\eqref{eq:GD_Hamiltonian}], which only emerges for non-zero detunings ($\delta\neq0$). Thus, for the parameter regimes in which photon counting and homodyne detection are optimal, critically enhanced sensitivity is not observed (see Fig.~\ref{Fig3}).\\

\noindent \textbf{Discussion.---}
We derived conditions for open many-body systems under which the system-environment QFI is saturated by ideal photon counting and homodyne detection. This allows us to understand which parameters can be efficiently sensed and which terms in the Hamiltonian can be detrimental for such task. Importantly, we have shown that the saturation of the ultimate bound is not restricted to models with a ``classical'' scaling of the sensitivity but it can also be observed in the presence of quantum many-body enhancement. Our insights immediately provide guidance on how to construct quantum many-body models that saturate the system-environment QFI via photon counting and homodyne detection. In this sense, our results pave the way for experimental realizations and help to design open systems with quantum enhanced sensing via continuous monitoring.

The framework presented here does not only apply to spin-boson models but generically to any kind of open quantum system which can be treated within the input-output formalism of Eq.~\eqref{eq:pure_sys_env}, such as the \textit{boundary time-crystal} model \cite{iemini_boundary_2018} (see SM~\cite{SM}). Non-Markovian dynamics can also be naturally incorporated, by singling out the bath modes which have the strongest coupling to the system. Integrating out the remaining modes then gives rise to models as those investigated in this manuscript (see Ref.~\cite{tamascelli_nonperturbative_2018}).
Future directions of research involve deriving bounds on the deviation from the saturation for models outside the class and generalising the presented results to the case with unmonitored decay channels.
This is especially interesting in connection with recent trapped-ion experiments where small deviations from the ideal sensing design and detection inefficiencies are unavoidable.\\

\noindent The code and data that support the findings of this work are available on Zenodo~\cite{mattes_2025_zenodo}.\\

\noindent \textbf{Acknowledgments.---} We would like to thank M. Gu\c{t}\u{a} for interesting discussions on the saturation of the system-environment QFI, and K. M\o{}lmer for interesting comments on optimality of homodyne detection. We acknowledge support from the Deutsche Forschungsgemeinschaft (DFG, German Research Foundation) through the Walter Benjamin programme, Grant No. 519847240 and the Research Unit FOR 5413/1, Grant No.~465199066. FC~is indebted to the Baden-W\"urttemberg Stiftung for the financial support of this research project by the Eliteprogramme for Postdocs. We acknowledge support by the state of Baden-Württemberg through bwHPC and the German Research Foundation (DFG) through grant no INST 40/575-1 FUGG (JUSTUS 2 cluster). We acknowledge support from the Leverhulme Trust (Grant No. RPG-2024-112). This work was supported by the QuantERA II programme (project CoQuaDis, DFG Grant No. 532763411) that has received funding from the EU H2020 research and innovation programme under GA No. 101017733. This work is also supported by the ERC grant OPEN-2QS (Grant No. 101164443, https://doi.org/10.3030/101164443).

\bibliography{refs.bib}
\newpage

\setcounter{equation}{0}
\setcounter{figure}{0}
\setcounter{table}{0}
\makeatletter
\renewcommand{\theequation}{S\arabic{equation}}
\renewcommand{\thefigure}{S\arabic{figure}}
\makeatletter

\onecolumngrid
\newpage

\setcounter{page}{1}
\begin{center}
{\Large SUPPLEMENTAL  MATERIAL}
\end{center}
\begin{center}
\vspace{0.8cm}
{\Large Designing open quantum systems for enabling quantum enhanced sensing through classical measurements}
\end{center}
\begin{center}
Robert Mattes,$^{1}$ Albert Cabot,$^{1}$ Federico Carollo,$^{2}$ and Igor Lesanovsky$^{1,3}$
\end{center}
\begin{center}
$^1${\em Institut f\"ur Theoretische Physik and Center for Integrated Quantum Science and Technology, Universit\"at T\"ubingen,}\\
{\em Auf der Morgenstelle 14, 72076 T\"ubingen, Germany}\\
$^2${\em Centre for Fluid and Complex Systems, Coventry University, Coventry, CV1 2TT, United Kingdom}\\$^3${\em School of Physics and Astronomy and Centre for the Mathematics}\\
{\em and Theoretical Physics of Quantum Non-Equilibrium Systems,}\\
{\em  The University of Nottingham, Nottingham, NG7 2RD, United Kingdom}
\end{center}
\section{I. Generalized Dicke model}
\noindent For the interaction given in Eq.~\eqref{eq:GD_Hamiltonian} we recover the well-known open Dicke model~\cite{genway_generalized_2014,boneberg_quantum_2022}. In the thermodynamic limit we find for the mean-field operators $\underset{N\to\infty}{\lim} \langle S_c/S \rangle= m_c$ and $\underset{N\to\infty}{\lim} \langle a/\sqrt{S} \rangle = \alpha$ the following Heisenberg equations of motion
$$
\dot{m}_x = -\sqrt{2}\lambda m_p m_y\, , \,\, \dot{m}_y = -\Omega m_z +\sqrt{2}\lambda m_p m_x\, , \,\,\dot{m}_z = \Omega m_y\, , \,\, \dot{m}_q = \delta m_p + \sqrt{2}\lambda m_z - \frac{\kappa}{2}m_q\, , \,\,\dot{m}_p = -\delta m_q- \frac{\kappa}{2}m_p\, ,
$$
where $m_p = (\alpha + \alpha^*)/\sqrt{2}$ and $m_q=i(\alpha - \alpha^*)/\sqrt{2}$. Since the magnetization is conserved for the considered model we find with $\tilde{m}_z = \pm \sqrt{1 - (\tilde{m}_x)^2}$ the critical coupling strength $
\lambda_c = \sqrt{(\delta^2+(\kappa/2)^2)\Omega/2\delta}\, .
$
In analogy to Ref.~\cite{boneberg_quantum_2022} we find that the latter system of equations features for $\lambda < \lambda_c$ only one stable stationary solution given by
$$
    \tilde{m}_x = -1\, , \quad \tilde{m}_y = \tilde{m}_z = \tilde{m}_q = \tilde{m}_p = 0 \, ,
$$
and for $\lambda > \lambda_c$ two stable stationary states
$$
    \tilde{m}_x = -\frac{\lambda_c^2}{\lambda^2} \, , \quad \tilde{m}_y = 0 \, , \quad  \tilde{m}_z = \pm \sqrt{1 - (\tilde{m}_x)^2 }\, ,\quad \tilde{m}_q = \frac{\kappa \lambda\tilde{m}_z}{\sqrt{2}(\delta^2+(\kappa/2)^2)}\, ,\quad  \tilde{m}_p = -\frac{\sqrt{2}\lambda\delta\tilde{m}_z}{(\delta^2+(\kappa/2)^2)}\, .
$$\\

\begin{figure}[t]
    \centering
    \includegraphics[width=0.8\textwidth]{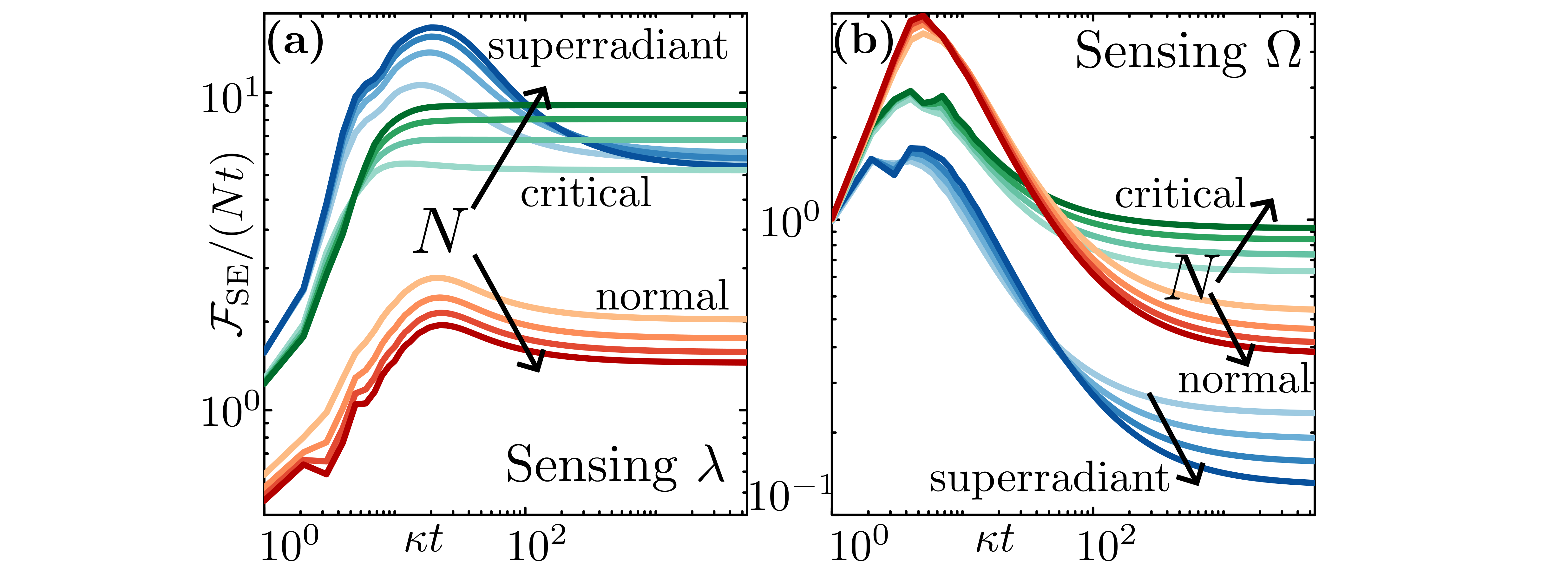}
    \caption{{\bf Many-body enhanced sensitivity in the generalized Dicke model.} Time-evolution of the system-environment QFI for $\Delta=0$, $\delta/\kappa=1$, $N=5,7,9,11$ and considering the interaction given in Eq.~\eqref{eq:GD_Hamiltonian}. (a) Sensitivity for estimating $\lambda/\kappa=1$ in the superradiant $\Omega/\kappa=0.4$ (blue), critical $\Omega/\kappa=1$ (green) and normal $\Omega/\kappa=2$ (red) regime. (b) Sensitivity for estimating $\Omega/\kappa=1$ in the normal $\lambda/\kappa=0.7$ (red), critical $\lambda/\kappa=1$ (green) and superradiant $\lambda/\kappa=1.3$ (blue) regime. The system is initially in the state $\ket{\psi(0)} = \ket{S,S}\otimes\ket{0}$.}
    \label{fig:SM1}
\end{figure}

\noindent In the following we will firstly discuss general properties of the system-environment QFI and secondly illustrate them for the generalized Dicke coupling~\eqref{eq:GD_Hamiltonian}.
While the state of the environment is generally a complicated object [cf. Eq.~\eqref{eq:pure_sys_env}], the system-environment QFI given in Eq.~\eqref{eq:ultimate_Fisher} can be, in principle, computed through the relation~\cite{gammelmark_fisher_2014,macieszczak_dynamical_2016}
\begin{equation}\label{eq:Fisher_deformed}
    \mathcal{F}_{\mathrm{SE}}(\eta, t) = 4 \partial_{\eta_1} \partial_{\eta_2} \log\left(\tr\left(\rho_{\eta_{12}}(t)\right)\right)\Big|_{\eta_1=\eta_2=\eta}\, . 
\end{equation}
Here, $\rho_{\eta_{12}}$ is an auxiliary operator evolving according to the two-sided master equation 
\begin{equation}\label{eq:Master_deformed}
    \frac{\mathrm{d}}{\mathrm{d} t} \rho_{\eta_{12}}= -iH(\eta_1)\rho_{\eta_{12}}+ i\rho_{\eta_{12}}H(\eta_2) + \mathcal{D}[\rho_{\eta_{12}}]\, ,
\end{equation}
with initial condition $|\psi(0)\rangle\langle\psi(0)|$.
We focus on the behavior of the system-environment QFI [cf. Eqs.~\eqref{eq:ultimate_Fisher} and \eqref{eq:Master_deformed}] for finite system sizes $N$, which reflects the situation in experiments~\cite{dogra_dissipation-induced_2019,Gilmore2021,sun_quantum_2025}. In the long-time limit, the reduced state of the system is associated with the unique finite-$N$ stationary state of the quantum master equation and a spectral decomposition 
of Eq.~\eqref{eq:Master_deformed} reveals an asymptotic linear increase of the QFI with time~\cite{gammelmark_fisher_2014,macieszczak_dynamical_2016}. Since the initial state is different from the stationary one, we find a transient regime characterized by a super-linear scaling of the system-environment QFI in time, which eventually approaches a linear one. The latter behavior is observed for both spin-boson couplings considered here and is illustrated in Figs.~(\ref{Fig1})-(\ref{fig:SM1}). Fig.~(\ref{fig:SM1}) further illustrates a superlinear scaling of the system-environement QFI with system size $N$ after the transient and close to the critical regime.

\section{II. Analytical values of the QFI for model I in the stationary phase}
\noindent Analog to Ref.~\cite{cabot_continuous_2024} we perform a Holstein-Primakoff transformation on the spins, where we assume that the resulting bosonic mode ($[b,b^\dagger]=1$) is in a large displaced state around which we want to study the fluctuations. For the cavity mode we assume the same such that $m_a^N = \alpha+a/\sqrt{S}$.\\

\noindent For the model considered in Eq.~\eqref{eq:TC_Hamiltonian} we have
\begin{align*}
  \partial_t \rho = &-iS\frac{\Omega}{2} \left[m_-^N  + m_+^N, \rho \right] -i\lambda S \left[m_a^N m_+^N + m_{a^\dagger}^N m_-^N,\rho\right]+\kappa S (m_a^N \rho m_{a^\dagger}^N-\frac{1}{2}(\{m_{a^\dagger}^N m_a^N, \rho\}))\, .
\end{align*}
Expanding in orders of $S$ leads to the first non-vanishing term
\begin{align*}
  &-i\sqrt{S} \frac{\Omega}{2} \left[m_{-,1}^N  + m_{+,1}^N, \rho \right] -i\sqrt{S}\lambda  \left[m_{a,0}^N m_{+,1}^N + m_{a^\dagger,0}^N m_{-,1}^N,\rho\right]\\\
  &-i \sqrt{S} \lambda\left[m_{a,1}^N m_{+,0}^N + m_{a^\dagger,1}^N m_{-,0}^N,\rho\right] +\frac{\sqrt{S}\kappa }{2} (m_{a,0}^N [\rho,m_{a^\dagger,1}^N]+ m_{a^\dagger,0}^N[m_{a,1}^N,\rho])\, ,
\end{align*}
and the stationary values
$$
  m_{a,0}^N  = m_{a^\dagger,0}^N = -\Omega/2\lambda\, ,\quad m_{-,0}^N = \kappa \Omega/4i\lambda^2\, , \quad m_{+,0}^N = -\kappa \Omega/4i\lambda^2\, .
$$
The next higher term has a vanishing contribution in $S$ and for the stationary state we find
\begin{align*}
  0&=-i\lambda \left(\alpha\ket{\alpha}\bra{\alpha}\otimes m_{+,1}^N\ket{\beta}\bra{\beta}  + m_{a^\dagger,1}^N\ket{\alpha}\bra{\alpha}\otimes m_{-,1}^N\ket{\beta}\bra{\beta}\right) \\\
  &+ i\lambda \left(\ket{\alpha}\bra{\alpha}m_{a,1}^N\otimes \ket{\beta}\bra{\beta}m_{+,1}^N  + \ket{\alpha}\bra{\alpha}\alpha^* \otimes \ket{\beta}\bra{\beta}m_{-,1}^N\right)\\\
  &+\kappa (|\alpha|^2\ket{\alpha}\bra{\alpha}-\frac{1}{2}(m_{a^\dagger,1}^N \alpha \ket{\alpha}\bra{\alpha} +   \ket{\alpha}\bra{\alpha} \alpha^* m_{a,1}^N ))\otimes \ket{\beta}\bra{\beta}\, ,
\end{align*}
where we used $m_{a,1}^N\ket{\alpha} = a\ket{\alpha} = \alpha\ket{\alpha}$. A solution of the equation above is given by
$$
  m_{a,1}^N\ket{\alpha} = a\ket{\alpha} = 0\, ,\quad m_{-,1}^N\ket{\beta}= 0\, .
$$
While the latter equation is fulfilled for a vacuum state for the cavity the spin subsystem is in a squeezed state, which corresponds to the results presented in Ref.~\cite{Mattes2023}.
With the eigenvalue of the deformed master equation $\mu$
we can provide an analytical expression for the Fisher information in the stationary regime. Expanding in orders of $S$ together with the stationary state of the fluctuations $\rho_0=\ket{E_0}\bra{E_0}$ we find 
$$
  \mu(\Omega_1,\Omega_2) = \kappa S [ \frac{\Omega_1\Omega_2}{4\lambda^2} - \frac{\Omega_1^2}{8\lambda^2}  - \frac{\Omega_2^2}{8\lambda^2} ] \quad \Leftrightarrow \frac{\mathcal{F}_{\mathrm{SE}}}{t} = \frac{\kappa S}{\lambda^2}\, .
$$
Notice that the QFI displays an additional factor $N$ with respect to the one reported in Ref. \cite{cabot_continuous_2024}. This is due to the fact that in the present work we rescale the interaction strength in Eq.~\eqref{eq:TC_Hamiltonian} by $N$, while in Ref. \cite{cabot_continuous_2024} no rescaling of the parameters is considered. A more detailed discussion on how rescaling system parameters can lead to different scaling behavior of the QFI can be found in \cite{lee_timescales_2025}. In a similar way it is possible to provide a bound on the sensitivity for estimating $\lambda$
$$
  \mu(\lambda_1,\lambda_2) = \kappa S [ \frac{\Omega^2}{4\lambda_1 \lambda_2} - \frac{\Omega^2}{8\lambda_1^2}  - \frac{\Omega^2}{8\lambda_2^2} ] \quad \Leftrightarrow \frac{\mathcal{F}_{\mathrm{SE}}}{t} = \frac{\kappa \Omega^2 S}{\lambda^4}\, .
$$

\section{III. Saturation of the bound via photon counting}\label{sec:sat_photo}

\subsection{Conditions for saturation of the system-environment QFI}

\noindent The QFI of the combined system and environment state $\ket{\Psi(t)}$ for a general parameter $\eta$ [cf. Eq.~\eqref{eq:pure_sys_env}] is given by 
\begin{equation}\label{eq:ultimate_Fisher_SM}
    \mathcal{F}_{\mathrm{SE}}(\eta,t) = 4\left( \braket{\partial_\eta \Psi(t)|\partial_\eta \Psi(t)} + (\braket{\partial_\eta \Psi(t)|\Psi(t)})^2\right)\, .
\end{equation}
The last term of this expression reads
\begin{align*}
    \braket{\Psi(t)|\partial_\eta\Psi(t)} &= \sum_{{\bf i_t}} \sum_{j=1}^M \braket{\psi(0)| K_{i_1}^{\dagger} \dots K_{i_M}^{\dagger} K_{i_M} \dots (\partial_\eta K_{i_j})\dots K_{i_1} |\psi(0)}\\\
    &= \sum_{j=1}^M \sum_{{\bf i_{t'}}} \braket{\psi(0)| K_{i_1}^{\dagger} \dots K_{i_j}^{\dagger} (\partial_\eta K_{i_j})\dots K_{i_1} |\psi(0)}\, ,
\end{align*}
with ${\bf i_{t'}} = \{i_1,\dots,i_j\}$. In the last step we used that for the Kraus operators $\sum_{\{i_{m}\}} K_{i_{m}}^{\dagger}K_{i_{m}} = \mathbf{1}$.
We assume that the parameter to be determined is contained within the Hamiltonian and thus we find that $\partial_\eta K_{1_{m}}= 0$ [cf. Eq.~\eqref{eq:Kraus_ops}]. However, the presented framework can also be used to derive conditions for parameters associated with the dissipator given in Eq.~\eqref{eq:Dissipator}. The only contributing term is therefore given by
\begin{equation}
    \braket{\Psi(t)|\partial_\eta\Psi(t)} = \sum_{j=1}^N \sum_{{\bf i_{t'-\Delta t}}} \braket{\psi(0)| K_{i_1}^{\dagger} \dots K_{i_{j-1}}^{\dagger} e^{\left(i\left(H+\frac{i}{2}L^\dagger L\right)\Delta t\right)} (-i\partial_\eta H\Delta t) e^{\left(-i\left(H-\frac{i}{2}L^\dagger L\right)\Delta t\right)} K_{i_{j-1}}\dots K_{i_1} |\psi(0)}\, ,
\end{equation}
Since the summation over all possible trajectories restores the notion of the full quantum state we find that the second term of Eq.~\eqref{eq:ultimate_Fisher_SM} vanishes if  the expectation value of the operator associated to the parameter is zero for all times, i.e. $\text{Tr}\big[\partial_\eta H \rho (t)\big]=0$ where $\rho(t)$ is the reduced state of the system only. Notice that this could also be enforced by use of a global time-dependent phase, see discussion in Sec.~VII.\\

\noindent On the level of quantum trajectories the classical Fisher information associated to the photon counting record is given by
\begin{equation}
    \mathcal{I}_{\mathrm{E}}^{(c)}(\eta,t) = \sum_{\bf{i_t}} \braket{\tilde{\psi}_{\bf{i_t}}|\tilde{\psi}_{\bf{i_t}}} \left(\braket{\psi_{\bf{i_t}}|\phi_{\bf{i_t}}} + \braket{\phi_{\bf{i_t}}|\psi_{\bf{i_t}}} \right)^2\, ,
\end{equation}
where $\ket{\tilde{\psi}_{\bf{i_t}}}$ is the unnormalized conditional state and thus provides the probability for a certain trajectory. The vector $\ket{\phi_{\bf{i_t}}}$ is defined as $\ket{\phi_{\bf{i_t}}}= \ket{\partial_\eta\tilde{\psi_{\bf{i_t}}}}/\braket{\tilde{\psi_{\bf{i_t}}}|\tilde{\psi_{\bf{i_t}}}}$, and it evolves from time $t'=j\Delta t$ to time $t'+\Delta t$ according to~\cite{albarelli_restoring_2018}
$$
    \ket{\phi_{{\bf i_{t'+\Delta t}}}} = \frac{(\partial_\eta K_{i_{j+1}})\ket{\psi_{{\bf i_{t'}}}} + K_{i_{j+1}}\ket{\phi_{{\bf i_{t'}}}}}{\sqrt{\braket{\psi_{{\bf i_{t'}}}|K_{i_{j+1}}^{\dagger} K_{i_{j+1}}|\psi_{{\bf i_{t'}}}}}}\, ,
$$
where $\ket{\psi_{\bf i_{t'}}}$ is the normalized conditional state associated to the quantum trajectory ${\bf i_{t'}}$. For the moment we assume $A_{\bf i_{t}}=\braket{\psi_{\bf i_{t}}|\phi_{\bf i_{t}}}=\braket{\phi_{\bf i_{t}}|\psi_{\bf i_{t}}}$, however we will prove later on that this is the case for the same class of models and states for which the second term in Eq.~\eqref{eq:ultimate_Fisher_SM} is equal to zero. With the latter result the classical Fisher information simplifies to 
$
    \mathcal{I}_{\mathrm{E}}^{(c)}(\eta,t) = 4 \sum_{\bf i_{t}} \braket{\tilde{\psi}_{\bf i_{t}}|\tilde{\psi}_{\bf i_{t}}} A_{\bf i_{t}}^2\, .
$
The average QFI of the states conditional on the trajectories is~\cite{albarelli_restoring_2018}
\begin{equation}\label{eq:FI_traj}
    \mathcal{F}_{\mathrm{S}}^{(c)}(\eta,t) =4 \sum_{\bf i_{t}} \braket{\tilde{\psi}_{\bf i_{t}}|\tilde{\psi}_{\bf i_{t}}} \left( \braket{\partial_\eta \psi_{\bf i_{t}}|\partial_\eta \psi_{\bf i_{t}}} + (\braket{\partial_\eta \psi_{\bf i_{t}}|\psi_{\bf i_{t}}})^2\right) = 4 \sum_{\bf i_{t}}\braket{\tilde{\psi}_{\bf i_{t}}|\tilde{\psi}_{\bf i_{t}}} \left( \braket{\phi_{\bf i_{t}}|\phi_{\bf i_{t}}} - A_{\bf i_{t}}^2\right)\, , 
\end{equation}
where we used that $\ket{\partial_\eta \psi_{\bf i_{t}}} = \ket{\phi_{\bf i_{t}}} - A_{\bf i_{t}} \ket{\psi_{\bf i_{t}}}$. Thus, we have 
\begin{equation}\label{eq:total_Fisher_traj}
    \mathcal{F}_{\mathrm{SE}}^{(c)}(\eta,t) = \mathcal{I}_{\mathrm{E}}^{(c)}(\eta,t) + \mathcal{F}_{\mathrm{S}}^{(c)}(\eta,t) = 4 \sum_{\bf i_{t}} \braket{\tilde{\psi}_{\bf i_{t}}|\tilde{\psi}_{\bf i_{t}}} \braket{\phi_{\bf i_{t}}|\phi_{\bf i_{t}}} = 4 \sum_{\bf i_{t}} \left| \left| \sum_{j=1}^M K_{i_M} \dots (\partial_\eta K_{i_j})\dots K_{i_1} \ket{\psi(0)}\right| \right|^2\, .
\end{equation}
Comparing it with Eq.~\eqref{eq:ultimate_Fisher_SM} reveals that system-environment QFI is saturated for the conditions given in Eq.~\eqref{eq:conditions_saturation}.\\

\subsection{Initial condition not belonging to the class of saturating states}

\noindent As we will see, there is a family of models for which the dynamics spontaneously brings the system state within the class of states saturating the bound both at the master equation level and at the level of single quantum trajectories. Therefore, we find that, for any initial condition and after a sufficiently large but finite time, both the conditional state of the system and its average state enter the class of saturating states, remaining in it from there on.

Denoting the last time step at which the system is not within the class by  $t^* = l\Delta t$, then the long-time QFI rate will be saturated by the one of the trajectories. This follows from the QFI of the combined system-environment state at time $t= M\Delta t>t^*$
$$
    \mathcal{F}_{\mathrm{SE}}(\eta,t) = 4\left( \braket{\partial_\eta \Psi(M\Delta t)|\partial_\eta \Psi(M\Delta t)} + (\braket{\partial_\eta \Psi(l\Delta t)|\Psi(l\Delta t)})^2\right)\, .
$$
For the QFI of the trajectories we make use of the fact that, for $t>t^*$, $\braket{\psi_{{\bf i_t}}|\phi_{{\bf i_t}}} = A_{\bf i_{t}}+iB_{\bf i_{t^*}}$. Even though the time step $l$ and therefore $B_{\bf i_{t^*}}$ depends on the specific trajectory ${\bf i_{t^*}}=\{i_1,\dots,i_l\}$, we assume that the system enters the target class for all trajectories within a finite time. While the classical Fisher information of the trajectories is given by the same expression as before, i.e.
$
    \mathcal{I}_{\mathrm{E}}^{(c)}(\eta,t) = 4 \sum_{\bf i_{t}} \braket{\tilde{\psi}_{\bf i_{t}}|\tilde{\psi}_{\bf i_{t}}} A_{\bf i_{t}}^2\, ,
$
we find for the QFI of the trajectories [cf. Eq.~\eqref{eq:FI_traj}] additional terms stemming from the non-vanishing imaginary value $B_{\bf i_{t^*}}$. The resulting expression in general splits up into two parts
$$
    \mathcal{F}_{\mathrm{SE}}^{(c)}(\eta,t) = 4 \sum_{\bf i_{t}} \braket{\tilde{\psi}_{\bf i_{t}}|\tilde{\psi}_{\bf i_{t}}} \braket{\phi_{\bf i_{t}}|\phi_{\bf i_{t}}} - \sum_{\bf i_{t^*}} \braket{\tilde{\psi}_{\bf i_{t^*}}|\tilde{\psi}_{\bf i_{t^*}}} B_{\bf i_{t^*}}^2\, , 
$$
and while the contribution of the first term increases with time, the second one remains constant for all $M>l$ (see examples below). Therefore, we find saturation of the QFI rate in the long-time limit [see Fig.~\ref{fig:SM2}]
$$
    \underset{t\to\infty}{\lim} \frac{(\mathcal{F}_{\mathrm{SE}}^{(c)}(\eta,t) - \mathcal{F}_{\mathrm{SE}}(\eta,t))}{t} = 0.
$$

\section{IV. Class of models saturating the bound with photon counting}
\noindent Firstly, we show that conditions (I) and (II) are fulfilled for a class of states appearing in the considered models. Secondly, we show that while for model I states outside the class eventually evolve into states of the class and remain there, this is not the case for model II.\\

\noindent We are considering generic open spin-boson models [cf. Fig.~\ref{Fig1}], where the spin system couples collectively to the bosonic mode. The spin degree of freedom can be thus described within the Dicke state representation $\ket{S, M_z}$, with the angular momentum $S$ and the magentization in $z$-direction $M_z=-S, -S+1,\dots, S$~\cite{carmichael_analytical_1980}. Together with the bosonic occupation number $n$ we have states of the form
$$
    \ket{\psi} = \sum_{M_z = -S}^S \sum_{n=0}^\infty  C_{M_z,n} (\ket{S, M_z}\otimes\ket{n})\, .
$$
The operator $\partial_\eta H$ is for the class of models considered in this work in general a combination of bosonic $\{a,a^\dagger,\mathbf{I}_\textbf{B}\}$ and collective spin operators $\{S_{\pm},S_z,\mathbf{I}_{\textbf{S}}\}$. The action of these basic operators on the general state above illustrate which kind of states are the ones for which conditions (I) and (II) are fulfilled. The collective lowering operator $S_-$ acts on the spin-subspace and we find
$$
    S_- \sum_{M_z = -S}^S\sum_{n=0}^\infty C_{M_z,n}(\ket{S, M_z}\otimes\ket{n})  =  \sum_{M_z = -S}^{S-1}\sum_{n=0}^\infty \sqrt{(S+M_z+1)(S-M_z)} C_{M_z+1,n} (\ket{S, M_z}\otimes\ket{n})\, .
$$
The action of the annihilation operator on such a state is 
$$
    a \sum_{M_z = -S}^S\sum_{n=0}^\infty C_{M_z,n}(\ket{S, M_z}\otimes\ket{n}) = \sum_{M_z = -S}^{S}\sum_{n=0}^\infty \sqrt{n+1} C_{M_z,n+1} (\ket{S, M_z}\otimes\ket{n})\, .
$$
With these results we can identify a class of states satisfying conditions (I) and (II), whereby the class depends on the parameter $\eta$ and the specific model.\\

\noindent \textbf{Model I.--} We consider a dissipative version of the Tavis-Cummings model with
\begin{equation}\label{eq:TC_SM}
    H = \frac{\Omega}{2}\left( S_-+S_+\right) +\frac{\lambda}{\sqrt{S}}(aS_+ + a^\dagger S_-)
\end{equation}
and photon losses at rate $\kappa$ [see Eq.~\eqref{eq:Dissipator}]. 
For states with the property
\begin{equation}\label{eq:condition_coefficents_TC_SM}
    C_{M_z,n}^* \cdot C_{M_z+1,n} \in \mathbb{I}\, \quad \mathrm{and} \quad C_{M_z,n}^* \cdot C_{M_z,n+1} \in \mathbb{R}\,,
\end{equation}
we find a purely imaginary expectation value for $S_-$ and $a^\dagger S_-$, which leads to a vanishing expectation value of $S_x$ and $(aS_+ + a^\dagger S_-)$. Condition (I) is fulfilled for $\partial_\Omega H = S_x$ or $\partial_\lambda H = (aS_+ + a^\dagger S_-)/\sqrt{S}$ at all times if the Kraus operators do not change the property of the state given in Eq.~\eqref{eq:condition_coefficents_TC_SM}. This can be seen from the expansion of the Kraus operators to first order in $\Delta t $
$$
    K_0= \mathbf{1}-i\left(\frac{\Omega}{2} (S_- + S_+) + \frac{\lambda}{\sqrt{S}}(aS_++a^\dagger S_-)\right)\Delta t -\frac{\kappa}{2}a^\dagger a \Delta t\, , \quad K_1=\sqrt{\Delta t \kappa} a\, .
$$
While the collective spin operators $S_{\pm}$ change the order of the state the ($-i$) in front restores the symmetry such that it remains together with the identity within the class. The term associated with $\lambda$ illustrates that coefficients connected by bosonic creation or annihilation operators should have the same property, see the second condition on the coefficients given in Eq.~\eqref{eq:condition_coefficents_TC_SM}. A non-vanishing detuning, i.e. $\Delta,\delta>0$ would immediately destroy the property of the state since terms proportional to $-iS_z$ or $-ia^\dagger a$ make the coefficients to acquire a general complex phase. Notice also that if the initial state is $|\psi(0)\rangle=|S,M_z\rangle\otimes|n\rangle$, a single application of $K_0$ brings it to the class of states defined by Eq.~\eqref{eq:condition_coefficents_TC_SM}.

\begin{figure}
    \centering
    \includegraphics[width=\textwidth]{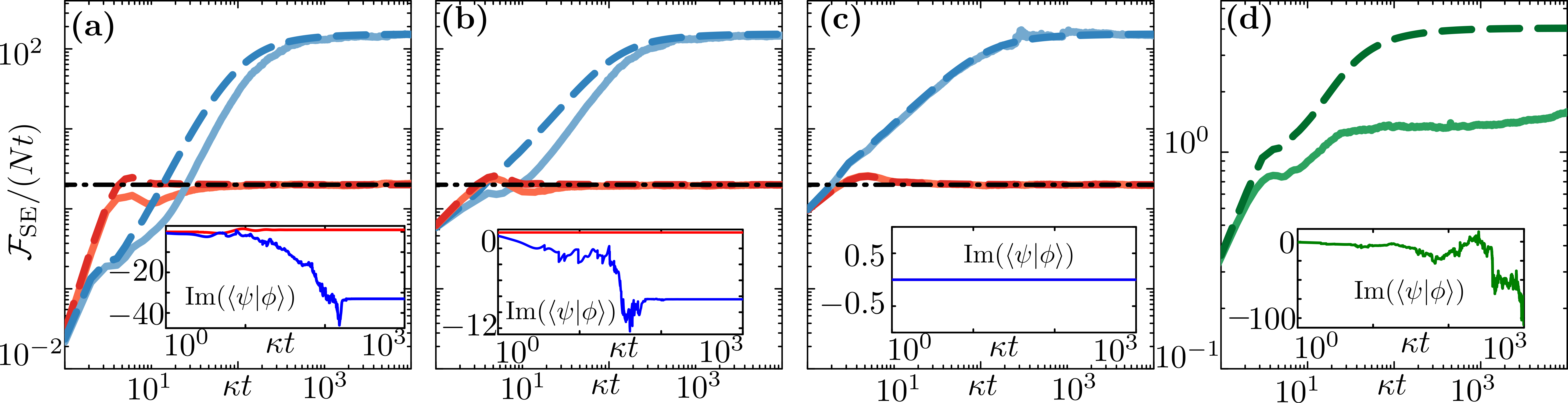}
    \caption{{\bf Saturation with photon counting for initial states outside the saturating class.} (a)-(c) System-environment QFI (dashed) and QFI of the photocounting trajectories (solid), averaged over $10^4$ trajectories, associated to sensing $\Omega$ for the model given in Eq.~\eqref{eq:TC_Hamiltonian}, with $\Omega/\kappa=0.1$ in the stationary (red) and $\Omega/\kappa=2$ in the time-crystal regime (blue) for $\lambda/\kappa=0.5$ and $N=5$. The dashed-dotted line is the analytical value in the stationary regime. The inset shows $\mathrm{Im}(\braket{\psi|\phi})$ for a single trajectory in the stationary (red) and the time-crystal (blue) regime. The initial state is hereby $\ket{S,\theta,\phi}$~\cite{carmichael_analytical_1980}, with (a) $(\theta,\phi) = (\pi/2,0)$, (b) $(\theta,\phi) = (\pi/4,\pi/8)$ and (c) $(\theta,\phi) = (\pi,0)$. (d) System-environment QFI (dashed) and QFI of the photocounting trajectories (solid), averaged over $10^4$ trajectories, associated to sensing $\lambda$ for the model given in Eq.~\eqref{eq:GD_Hamiltonian}, with $\Omega/\kappa=1$, $\lambda/\kappa=1$, $\delta/\kappa=0$, $N=5$ and the initial state $(\theta,\phi) = (\pi/2,0)$. The inset shows again the evolution of $\mathrm{Im}(\braket{\psi|\phi})$ for a single trajectory and latter parameters.}
    \label{fig:SM2}
\end{figure}

For condition (II) $\braket{\psi_{\bf i_{t}}|\phi_{\bf i_{t}}}=\braket{\phi_{\bf i_{t}}|\psi_{\bf i_{t}}}$ we consider a specific trajectory ${\bf i_{t}}=\{i_1,\dots,i_j,\dots,i_M\}$ such that 
$$
    \braket{\psi_{\bf i_{t}}|\phi_{\bf i_{t}}} = \sum_{j=1}^M \braket{\psi_{\bf i_{t'}}| K_{i_{j+1}}^{\dagger} \dots  K_{i_M}^{\dagger}  K_{i_M} \dots (-i\partial_\eta H\Delta t)|\psi_{\bf i_{t'}}}\, .
$$
Since the state at time $t'=j\Delta t$ fulfills the condition in Eq.~\eqref{eq:condition_coefficents_TC_SM} and $(-i\partial_\eta H)$ conserves this property we find that $\braket{\psi_{\bf i_{t}}|\phi_{\bf i_{t}}}\in\mathbb{R}$.
Further, we have numerically observed that if we start with an initial state which does not satisfy Eq.~\eqref{eq:condition_coefficents_TC_SM} the system will evolve into a state satisfying the latter, where the state picks up a global phase 
$$
    \ket{\psi} = e^{i\varphi}\sum_{M_z = -S}^S \sum_{n=0}^\infty  C_{M_z,n} (\ket{S, M_z}\otimes\ket{n})\, .
$$
Whereby the specific trajectory ${\bf i_{t^*}}=\{i_1,\dots,i_l\}$ in the transient defines the value of the phase.
The transient regime with $C_{M_z,n}^* \cdot C_{M_z+1,n} \in \mathbb{C}$ leads to a constant offset from the system-environment QFI. In the long-time limit we therefore find independent from the initial state the same scaling for the conditional and the system-environment QFI, as illustrated in Fig.~\ref{fig:SM2}.

\noindent Finally we remark that the \textit{boundary time-crystal} (BTC) \cite{iemini_boundary_2018} represents a special case of the model described in Eq.~\eqref{eq:TC_Hamiltonian}, and it is thus contained in the same class as Model I. In the limit $\kappa \gg \Omega, \lambda$ the bosonic mode can be adiabatically eliminated and the system evolves according to \cite{Mattes2023}
$$
    \dot{\rho} = -i\left[\frac{\Omega}{2}\left( S_-+S_+\right),\rho \right] + \frac{\gamma}{S}(S_-\rho S_+ -\frac{1}{2}\{S_+S_-,\rho\})\, .
$$
The condition for saturation of the system-environment QFI by the one of the trajectories for photon counting is therefore simply
\begin{equation}\label{eq:condition_coefficents_BTC_SM}
    C_{M_z}^* \cdot C_{M_z+1} \in \mathbb{I}\, , \quad \mathrm{where}\quad \ket{\psi} = \sum_{M_z = -S}^S  C_{M_z} \ket{S, M_z} .
\end{equation}
For the BTC we also find  that even for an initial state that does not fulfill the condition given in Eq.~\eqref{eq:condition_coefficents_BTC_SM} the system eventually enters the intended class of states. In fact, in Ref. \cite{Cabot2023} it was reported that for any initial condition, photocounting trajectories eventually reach the point $\langle \psi_{\bf i_{t^*+\Delta t}}| S_{x}|  \psi_{\bf i_{t^*+\Delta t}}\rangle =0$, remaining there for subsequent times. This is a signature of the conditional state spontaneously reaching the form given in Eq. (\ref{eq:condition_coefficents_BTC_SM}) through the action of the photocounting process.\\

\noindent \textbf{Model II.--}  For a generalized version of the open Dicke model with Hamiltonian
\begin{equation}\label{eq:GD_SM}
    H = \frac{\Omega}{2}\left( S_-+S_+\right) +\frac{\lambda}{\sqrt{S}}(a + a^\dagger)S_z
\end{equation}
and photon losses at rate $\kappa$ [see Eq.~\eqref{eq:Dissipator}], we find that conditions (I) and (II) are satisfied for states of the form
\begin{equation}\label{eq:condition_coefficents_GD_SM}
    C_{M_z,n}^* \cdot C_{M_z+1,n} \in \mathbb{I}\, \quad \mathrm{and} \quad C_{M_z,n}^* \cdot C_{M_z,n+1} \in \mathbb{I}\, .
\end{equation}
Analog to model I the term proportional to $\Omega$ defines the first condition and the interaction proportional to $\lambda$ dictates the second condition. Similarly to model I, one can show by application of the Kraus operators, that the dynamics maps states of this class into states of the same class. The coefficients loose their property, i.e. being purely real or imaginary, for a non-vanishing detuning and the QFI is therefore only saturated for $\delta=\Delta=0$.

Generally, for model II we do not observe the convergence into the class of states (\ref{eq:condition_coefficents_GD_SM}) if we start the dynamics with a state which lies outside the class. Thus, in order to saturate the QFI it is necessary to initialize the system in the proper initial condition.

\section{V. Saturation of the bound via homodyne detection}\label{sec:sat_homo}

\subsection{System-environment joint state in terms of the homodyne unravelling}

In this section, following the discrete-time version of input-output theory, we write down the system-environment joint state in terms of the homodyne Kraus operators. For the initial state, we consider the tensor product of the system in a pure state and all {\it time-bin} modes in the vacuum state~\cite{gammelmark_fisher_2014,yang_efficient_2023,cabot2025}:
$$
|\Psi(0)\rangle=|\psi(0)\rangle \otimes(\otimes_{j=1}^{\infty}|0_j\rangle)\, .   
$$
The time-bin modes are the instances of the environment with which the system interacts sequentially for a small time step $\Delta t$. Each time-bin mode is represented by a bosonic degree of freedom with creation and annihilation operators $b_j$, $b_j^\dagger$, with $[b_j,b_j^\dagger]=1$. The interaction of the system with time-bin mode $j$ is given by the following unitary~\cite{gammelmark_fisher_2014,yang_efficient_2023,cabot2025}:
$$
U_j=\exp\big[-iH\Delta t+\sqrt{\Delta t}(L b_j^\dagger-L^\dagger b_j)\big]\, ,    
$$
where $H$ is the system Hamiltonian and $L$ is the jump operator, where the phase $\Phi$ [cf. Eq.~\eqref{eq:Kraus_homodyne}] is absorbed into the jump operator. At time $t=M\Delta t$ the system-environment joint state can be written as:
\begin{equation}\label{eq:SE_state_general}
\begin{split}
|\Psi(M\Delta t)\rangle&= \prod_{j=1}^M U_j |\psi(0)\rangle \otimes(\otimes_{j=1}^{\infty}|0_j\rangle)\\
&= \prod_{j=1}^M\bigg[\mathbf{1}-iH\Delta t-\frac{L^\dagger L}{2}\Delta t+\sqrt{\Delta t}L(b_j+b_j^\dagger) +\mathcal{O}(\Delta t^{3/2})\bigg]|\psi(0)\rangle\otimes|0_1,\dots, 0_M\rangle\, .
\end{split}
\end{equation}
In the second line we have used the expansion up to order $\mathcal{O}(\Delta t^{3/2})$ of the unitary $U_j$ and we have obviated the time-bin modes that have not yet interacted with the system. Following  \cite{Wiseman2009}, we have additionally introduced the term $\propto L b_j$ in the second line as when applied to $|0_j\rangle$ is identically zero. If we now project Eq.~\eqref{eq:SE_state_general} on the Fock states of each time-bin mode, we would obtain the usual expression of the system-environment state as a superposition of all photocounting trajectories presented in the main text~\cite{gammelmark_fisher_2014,yang_efficient_2023,cabot2025}. Here, we proceed alternatively by projecting each unitary on the (continuous) position basis of the time-bin modes \cite{Wiseman2009}. We use the following eigenstates of the quadrature operator:
$$
\frac{1}{\sqrt{2}}(b_j^\dagger + b_j)|q_j\rangle=q_j|q_j\rangle\, ,\quad \mathbf{1}_j=\int_{-\infty}^\infty dq_j |q_j\rangle\langle q_j|,\quad  \langle 0_j|q_j\rangle=\frac{1}{\pi^{1/4}}\exp\bigg[-\frac{q_j^2}{2}\bigg]\, .    
$$
The system-environment joint state can be rewritten as:
$$
|\Psi(M\Delta t)\rangle=\frac{1}{\pi^{\frac{M}{4}}} \prod_{j=1}^M \int_{-\infty}^\infty dq_j e^{-\frac{q_j^2}{2}} \bigg[\mathbf{1}-iH\Delta t-\frac{L^\dagger L}{2}\Delta t+\sqrt{2\Delta t}L q_j +\mathcal{O}(\Delta t^{3/2})\bigg]|\psi(0)\rangle\otimes |q_1,\dots q_M\rangle\, .
$$
We now define the homodyne current at time $t_j=j\Delta t$ as:
$$
J_j=\sqrt{\frac{2}{\Delta t}}q_j\, ,   
$$
from which, we can identify the homodyne Kraus operators:
\begin{equation}\label{eq:Kraus_homodyne_SM}
K_{J_j}= \mathbf{1}-iH\Delta t-\frac{L^\dagger L}{2}\Delta t+ L J_j \Delta t+\mathcal{O}(\Delta t^{3/2})\, .
\end{equation}
We have chosen not to include the factor $ \langle 0_j|q_j\rangle$ in the definition of the Kraus operators. This is a valid choice as long as we take this factor into account as an integration measure when summing over all possible outcomes, or when computing the probabilities of the measurement outcomes \cite{Wiseman2009}. In this way, we find that the Kraus operators satisfy:
$$
\sqrt{\frac{\Delta t}{2\pi}}\int_{-\infty}^\infty   dJ_j e^{-\frac{\Delta t}{2}J_j^2}  K_{J_j}^\dagger K_{J_j}=\mathbf{1}_j+\mathcal{O}(\Delta t^{3/2})\, ,
$$
as expected. Therefore, the expression of the system-environment joint state in terms of the homodyne Kraus operators is given by 
\begin{equation}\label{eq:SE_state_homodyne}
|\Psi(M\Delta t)\rangle=\bigg(\frac{\Delta t}{2\pi}\bigg)^{\frac{M}{4}}  \int_{-\infty}^\infty dJ_M e^{-\frac{\Delta t}{4}J_M^2}\dots\int_{-\infty}^\infty dJ_1 e^{-\frac{\Delta t}{4}J_1^2}\,\,  K_{J_M}\dots K_{J_j}\dots K_{J_1}    |\psi(0)\rangle\otimes |J_1,\dots, J_M\rangle\, .
\end{equation}

\subsection{Homodyne quantum trajectories}

A single realization of the homodyne process of length $t=M\Delta t$ is characterized by the measured homodyne record. For convenience, we use the same symbol as for the photocount record, i.e. ${\bf i_t}$, but  we now define it as the vector containing the values of the homodyne current at each time step: ${\bf i_t}=\{J_1,\dots,J_M\}$, with $J_j\in\mathbb{R}$. The conditional state of the system for a given homodyne trajectory reads:
\begin{equation}\label{eq:homodyne_trajectory}
|\psi_{\bf i_t}\rangle=\frac{K_{J_M}\dots K_{J_j}\dots K_{J_1}|\psi(0)\rangle}{\sqrt{\text{Tr}\{K_{J_M}\dots K_{J_j}\dots K_{J_1}|\psi(0)\rangle\langle \psi(0)|K_{J_1}^\dagger\dots K_{J_j}^\dagger\dots K_{J_M}^\dagger \}}}\, .    
\end{equation}
The numerator of Eq.~\eqref{eq:homodyne_trajectory} defines the corresponding unnormalized system state $|\tilde{\psi}_{\bf i_t}\rangle$, which can be used, together with the integration measure, to compute the probability of observing such a trajectory. We can also use this to compute the probability of getting the outcome $J_j$ at each time step. In particular, using $t'=j\Delta t$, it reads:
$$
\text{Prob}(J_j)=dJ_j \sqrt{\frac{\Delta t}{2\pi}}e^{-\frac{\Delta t}{2}J_j^2}\langle \psi_{\bf i_{t'-\Delta t}}|K_{J_{j}}^\dagger  K_{J_{j}}|\psi_{\bf i_{t'-\Delta t}}\rangle\, .    
$$
Using the definition of the homodyne Kraus operators and keeping the dominant terms up to order $\mathcal{O}(\Delta t)$ \cite{Wiseman2009}, we arrive to: 
\begin{equation}\label{eq:homodyne_statistics}
\text{Prob}(J_j)=dJ_j \sqrt{\frac{\Delta t}{2\pi}} \exp\bigg[{-\frac{\Delta t}{2}\big(J_j-\langle \psi_{\bf i_{t'-\Delta t}}|(L+L^\dagger)|\psi_{\bf i_{t'-\Delta t}}\rangle}\big)^2  \bigg]+\mathcal{O}(\Delta t)\, .
\end{equation} 
In a single realization, $J_j$ can be regarded as a Gaussian random number with average $\langle \psi_{\bf i_{t'-\Delta t}}|(L+L^\dagger)|\psi_{\bf i_{t'-\Delta t}}\rangle$ and variance $1/\Delta t$. Therefore, Eq.~\eqref{eq:homodyne_statistics} describes the statistics of the homodyne current when monitoring the quadrature $(L+L^\dagger)$ \cite{Wiseman2009}.

\subsection{Conditions for saturation of the system-environment QFI}

The conditions (I) and (II) for saturation of the system-environment QFI apply also to the homodyne unravelling, in which case we just need to make use of the homodyne Kraus operators and the corresponding measured record ${\bf i_t}=\{J_1,\dots,J_M\}$. For completeness, here we rewrite the expression for the different QFIs in terms of the homodyne Kraus operators.

We begin with the system-environment QFI, and we assume that condition (I) holds. Then the QFI is proportional to $\langle \partial_\eta \Psi(t)|\partial_\eta \Psi(t)\rangle$. Assuming that the parameter of interest is encoded in the system Hamiltonian, using Eq.~\eqref{eq:SE_state_homodyne} and the orthogonality properties of the quadrature eigenstates, we arrive at:
\begin{equation}\label{eq:SE_QFI_homodyne}
\mathcal{F}_\mathrm{SE}(\eta,t)=4\bigg(\frac{\Delta t}{2\pi}\bigg)^{\frac{M}{2}}\int_{-\infty}^\infty  dJ_M e^{-\frac{\Delta t}{2}J_M^2}\dots\int_{-\infty}^\infty dJ_1  e^{-\frac{\Delta t}{2}J_1^2}\,\, \bigg\|\bigg[\sum_{j=1}^M K_{J_M}\dots(\partial_\eta K_{J_j})\dots K_{J_1}    \bigg]|\psi(0)\rangle \bigg\|^2\ .  
\end{equation}
This expression is analogous to the one obtained in terms of the photocunting quantum trajectories, but replacing the corresponding Kraus operators by the homodyne ones, and the sum over all possible outcomes by the respective (weighted) integrals.

The Fisher information for the homodyne trajectories, $\mathcal{F}^{(h)}_\mathrm{SE}(\eta,t)$, can be computed following the same general recipe as for the case of photon counting \cite{albarelli_restoring_2018}. In particular,  for the contribution of the homodyne record we use 
$$
|\phi_{\bf{i_{t'+\Delta t}}}\rangle=  \frac{\partial_\eta K_{J_{j+1}}|\phi_{\bf{i_{t'}}}\rangle+K_{J_{j+1}}|\psi_{\bf{i_{t'}}}\rangle}{\sqrt{\langle \psi_{\bf{i_{t'}}}|K^\dagger_{J_{j+1}}K_{J_{j+1}}|\psi_{\bf{i_{t'}}}\rangle}},  
$$
and define $A_{\bf i_t} + iB_{\bf i_t}=\langle \psi_{\bf i_t}|\phi_{\bf i_t}\rangle$. Then, we arrive at:
$$
\mathcal{I}_{\mathrm{E}}^{\mathrm{(h)}}(\eta, t)=4\bigg(\frac{\Delta t}{2\pi}\bigg)^{\frac{M}{2}}\int_{-\infty}^\infty  dJ_M e^{-\frac{\Delta t}{2}J_M^2}\dots\int_{-\infty}^\infty dJ_1  e^{-\frac{\Delta t}{2}J_1^2}\,\, \text{Tr}\{K_{J_M}\dots K_{J_1}|\psi(0)\rangle\langle \psi(0)|K_{J_1}^\dagger\dots K_{J_M}^\dagger \} \, A_{\bf i_t}^2.
$$
This expression is analogous to the one we obtained for $\mathcal{I}_{\mathrm{E}}^{\mathrm{(c)}}(\eta, t)$ but replacing the sum over all counting records by the corresponding (weighted) integrals over all possible values of the homodyne current. The second contribution comes from the average  QFI for the conditional system state to each homodyne trajectory. Weighting each possible trajectory by its corresponding probability of occurrence, we obtain:
\small
$$
\mathcal{F}_{\mathrm{S}}^{\mathrm{(h)}}(\eta, t)=4\bigg(\frac{\Delta t}{2\pi}\bigg)^{\frac{M}{2}}\int_{-\infty}^\infty  dJ_M e^{-\frac{\Delta t}{2}J_M^2}\dots\int_{-\infty}^\infty dJ_1  e^{-\frac{\Delta t}{2}J_1^2}\,\, \text{Tr}\{K_{J_M}\dots K_{J_1}|\psi(0)\rangle\langle \psi(0)|K_{J_1}^\dagger\dots K_{J_M}^\dagger \} \, \big[ \langle \partial_\eta \psi_{\bf i_t}|\partial_\eta \psi_{\bf i_t}\rangle +\langle \partial_\eta \psi_{\bf i_t}|\psi_{\bf i_t}\rangle^2 \big].
$$
\normalsize

If we now assume condition (II), then $B_{\bf i_t}=0$ for all trajectories.  The series of steps that allowed us to write 
$$
\braket{\partial_\eta \psi_{\bf i_{t}}|\partial_\eta \psi_{\bf i_{t}}} + (\braket{\partial_\eta \psi_{\bf i_{t}}|\psi_{\bf i_{t}}})^2=\braket{\phi_{\bf i_{t}}|\phi_{\bf i_{t}}} - A_{\bf i_{t}}^2
$$ 
do not actually depend on who are the Kraus operators \cite{albarelli_restoring_2018}. Therefore, we arrive at:
\begin{equation}\label{eq:QFI_homodyne}
\begin{split}
\mathcal{F}^{(h)}_\mathrm{SE}(\eta,t) &=4\bigg(\frac{\Delta t}{2\pi}\bigg)^{\frac{M}{2}}\int_{-\infty}^\infty  dJ_M e^{-\frac{\Delta t}{2}J_M^2}\dots\int_{-\infty}^\infty dJ_1  e^{-\frac{\Delta t}{2}J_1^2}\,\, \text{Tr}\{K_{J_M}\dots K_{J_1}|\psi(0)\rangle\langle \psi(0)|K_{J_1}^\dagger\dots K_{J_M}^\dagger \} \, \langle \phi_{\bf i_t}| \phi_{\bf i_t}\rangle\\ &=4\bigg(\frac{\Delta t}{2\pi}\bigg)^{\frac{M}{2}}\int_{-\infty}^\infty  dJ_M e^{-\frac{\Delta t}{2}J_M^2}\dots\int_{-\infty}^\infty dJ_1  e^{-\frac{\Delta t}{2}J_1^2}\,\, \bigg\|\bigg[\sum_{j=1}^M K_{J_M}\dots(\partial_\eta K_{J_j})\dots K_{J_1}    \bigg]|\psi(0)\rangle \bigg\|^2\, ,  
\end{split}
\end{equation}
which coincides with the expression of $\mathcal{F}_\mathrm{SE}(\eta,t)$.\\

\begin{figure}
    \centering
    \includegraphics[width=\textwidth]{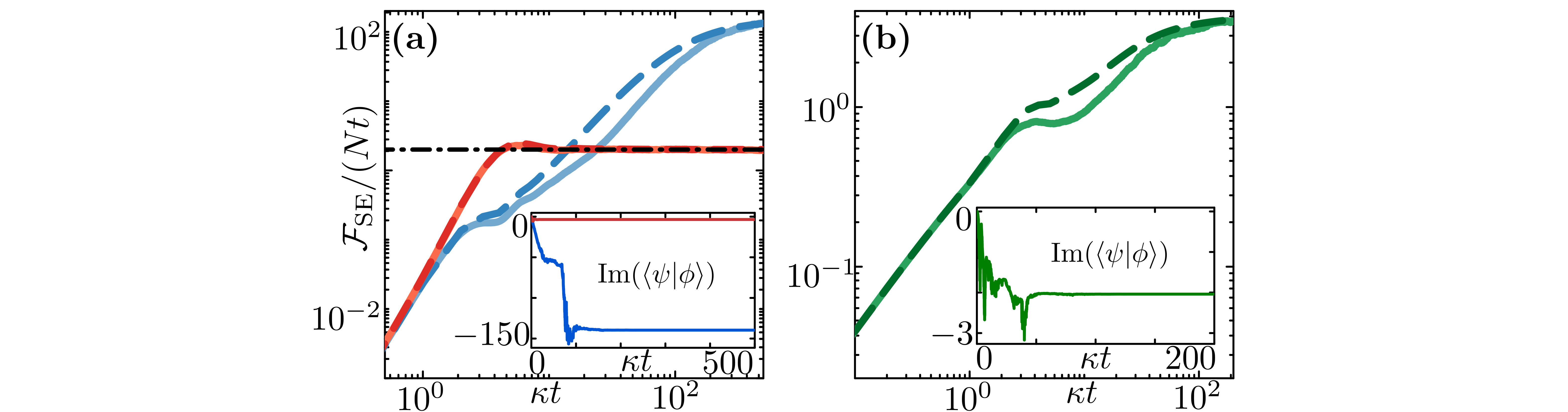}
    \caption{{\bf Saturation with homodyne detection for initial states outside the saturating class.} System-environment QFI (dashed) and QFI for homodyne detection (solid), averaged over $10^4$ trajectories for the initial state $\ket{S,\theta,\phi}$~\cite{carmichael_analytical_1980}, with $(\theta,\phi) = (\pi/2,0)$. (a) QFI associated with sensing $\Omega$ for the model given in Eq.~\eqref{eq:TC_Hamiltonian}, with $\Omega/\kappa=0.1$ in the stationary (red) and $\Omega/\kappa=2$ in the time-crystal regime (blue) for $\lambda/\kappa=0.5$ and $N=5$. The dashed-dotted line is the analytical value in the stationary regime. The inset shows $\mathrm{Im}(\braket{\psi|\phi})$ for a single trajectory in the stationary (red) and the time-crystal (blue) regime. (b) QFI associated with sensing $\lambda$ for the model given in Eq.~\eqref{eq:GD_Hamiltonian}, with $\Omega/\kappa=1$, $\lambda/\kappa=1$, $\delta/\kappa=0$ and $N=5$. The inset shows again the evolution of $\mathrm{Im}(\braket{\psi|\phi})$ for a single trajectory and latter parameters.}
    \label{fig:SM3}
\end{figure}

\noindent Analog to the photocounting process, it is possible to show that that for homodyne detection $$
    \underset{t\to\infty}{\lim} \frac{\left(\mathcal{F}_{\mathrm{SE}}^{(u)}(\eta,t) - \mathcal{F}_{\mathrm{SE}}(\eta,t)\right)}{t} = 0\, , 
$$
if we assume that the dynamics associated with the homodyne unravelling spontaneously brings the system state, within a finite time, into the class of states saturating the bound, see Fig.~\ref{fig:SM3}. 

\section{VI. Class of models saturating the bound with homodyne detection} 

For the homodyne unravelling an analysis analog to photon counting reveals the conditions for which one exhausts the system-environment QFI. The phase $\Phi$ of the local strong oscillator used for homodyne measurements is an additional degree of freedom and as we will illustrate in the following the optimal choice of the phase $\Phi$ explicitly depends on the considered model.\\

\noindent \textbf{Model I.--} For the dissipative version of the Tavis-Cummings model [cf. Eq.~\eqref{eq:TC_SM}] the saturation conditions \eqref{eq:conditions_saturation} are fulfilled for $\Omega$ and $\lambda$ for states of the same class~\eqref{eq:condition_coefficents_TC_SM} and $\Phi=0$. This can be seen from the corresponding Kraus operators
$$
    K_{J_j}= \mathbf{1}-i\left(\frac{\Omega}{2} (S_- + S_+) + \frac{\lambda}{\sqrt{S}}(aS_++a^\dagger S_-)\right)\Delta t -\frac{\kappa a^\dagger a}{2}\Delta t+ e^{i\Phi}\sqrt{\kappa} a\, \Delta t J_j\, .
$$
The second condition in Eq.~\eqref{eq:condition_coefficents_TC_SM} describes the relation between coefficients of the state connected by bosonic operators, i.e. the phase of the coefficient does not change. This property is conserved under the evolution with the latter Kraus operator for $\Phi=0$. For a non-zero value of the phase, the term proportional to the homodyne current introduces a phase between the coefficients $C_{M_z, n}$ and $C_{M_z, n-1}$, such that the system leaves the class of saturating states [see Fig.~\ref{Fig2}].

For the \textit{boundary time-crystal} homodyne detection with $\Phi=\pi/2$ is as well able to exhaust the system-environment QFI, which can be seen from the saturation condition for the states in Eq.~\eqref{eq:condition_coefficents_BTC_SM}.\\

\noindent \textbf{Model II.--} In contrast we find saturation for the generalized Dicke model given in Eq.~\eqref{eq:GD_SM} for states with $C_{M_z,n}^* \cdot C_{M_z,n+1} \in \mathbb{I}$, where these coefficients are associated to states connected by a bosonic creation operator. For a phase of $\Phi=\pi/2$, we stay within the class of saturating states. Our framework thus rigorously explains the results presented in Ref.~\cite{kiilerich_bayesian_2016} for a driven qubit with decay.

\section{VII. Connection of Eq.~(8) to optimality conditions for generic pure states}
\noindent We remark that the conditions [cf. Eq.~\eqref{eq:conditions_saturation}] derived in Sec. III. and V. for photon counting and homodyne detection to be optimal are, as we will show in the following, fundamentally related to the general necessary and sufficient conditions for a measurement to saturate the QFI of a generic pure state derived in Ref.~\cite{braunstein_statistical_1994}. In multipartite quantum systems, the quest for local optimal measurements is a subject of current research~\cite{zhou_saturating_2020,godley_adaptive_2023,yang_efficient_2023}. The results presented in the main text address this issue establishing that for a broad class of Markovian quantum dynamics such a local optimal measurement corresponds to the experimentally relevant continuous monitoring schemes of photon counting and homodyne detection.\\

\noindent Starting point of our considerations is again the pure system-environment state $\ket{\Psi(t)} = \sum_{\mathbf{i}_t}  \ket{\tilde{\psi}_{\mathbf{i}_t} }\otimes \ket{\mathbf{i}_t}$ and a set of generalized measurements $\{\Pi_{\mathbf{i}_t,j}\}$, with $\Pi_{\mathbf{i}_t,j} = M_{\mathbf{i}_t,j}\otimes\ket{\mathbf{i}_t}\bra{\mathbf{i}_t}$, where $\ket{\mathbf{i}_t}\bra{\mathbf{i}_t}$ is a projector onto a certain (continuous monitoring) trajectory $\mathbf{i}_t=\{i_1,\dots,i_M\}$, i.e. a classical measurement record of the emission. The latter form of the projector is as our derivation generic, and it will be specified by the continuous monitoring protocol performed on the output, e.g. homodyne or photocounting unravelling. For every trajectory $\mathbf{i}_t$ we have defined a set of \emph{system only} measurement operators $\{M_{\mathbf{i}_t,j}\}$ , with $\sum_j M_{\mathbf{i}_t,j}^\dagger M_{\mathbf{i}_t,j} = \mathbf{1}_S$, where $\mathbf{1}_S$ is the identity in the system's Hilbert space. Our goal is to show that if Conditions (I)-(II) [cf. Eq.~\eqref{eq:conditions_saturation}] are satisfied, this specific class of measurements (i.e., measurement operators being separable on system and output degrees of freedom, and output measurements corresponding to classical time-local measurements) is optimal for the considered class of open quantum systems. From Ref.~\cite{braunstein_statistical_1994} we can read the conditions for a measurement set to be optimal given a pure state. Introducing our measurement,  $\{\Pi_{\mathbf{i}_t,j}\}$, our pure state, $\ket{\Psi(t)}$, and the parameter of interest, $\eta$, the conditions that should be satisfied for all measurement outcomes are: \\
\begin{equation}\label{eq:sat_cond_gen_pure}
    \begin{split}    
        (1)\quad &\mathrm{Im}(\braket{\Psi|\Pi_{\mathbf{i}_t,j}^\dagger \Pi_{\mathbf{i}_t,j} |\partial_\eta \Psi^\perp} ) = 0\, ,\\
        (2)\quad &\Pi_{\mathbf{i}_t,j}[\ket{\Psi} - 2 \lambda_{\mathbf{i}_t,j} \ket{\partial_\eta \Psi^\perp}]  = 0\, , \quad \mathrm{where} \quad \lambda_{\mathbf{i}_t,j} = \frac{\braket{\Psi|\Pi_{\mathbf{i}_t,j}^\dagger \Pi_{\mathbf{i}_t,j} |\Psi}}{2\braket{\Psi|\Pi_{\mathbf{i}_t,j}^\dagger \Pi_{\mathbf{i}_t,j} |\partial_\eta \Psi^\perp}} \,.
    \end{split}
\end{equation}
Here we have used the definition: $ \ket{\partial_\eta \Psi^\perp}=\ket{\partial_\eta \Psi} - \ket{ \Psi}\braket{ \Psi|\partial_\eta \Psi}$. If these conditions are satisfied, the Fisher information of this measurement, $\mathcal{F}_{\mathrm{SE}}^{(u)} (\eta,t)$, saturates the QFI of the system-environment joint state, $\mathcal{F}_{\mathrm{SE}} (\eta,t)$.

In order to be able to recover the full QFI of the compound system-environment state, the measurements $M_{\mathbf{i}_t,j}$ acting on a conditional system state $\ket{\psi_{\mathbf{i}_t}}$ need to be optimal, as they will provide the contribution $\mathcal{F}_{\mathrm{S}}^{(u)}(\eta,t)$ to the total Fisher information of the specific unravelling $\mathcal{F}_{\mathrm{SE}}^{(u)} (\eta,t)$ [see Eq.~\eqref{eq:Fisher_conditional}]. Thus, we assume that $M_{\mathbf{i}_t,j}$ satisfy the conditions of optimality for estimating $\eta$ given the pure state $\ket{\psi_{\mathbf{i}_t}}$, i.e. \cite{braunstein_statistical_1994}:
\begin{equation}\label{eq:sat_cond_gen_pure_sys}
    \begin{split}    
        (1)\quad &\mathrm{Im}(\braket{\psi_{\mathbf{i}_t}|M_{\mathbf{i}_t,j}^\dagger M_{\mathbf{i}_t,j} |\partial_\eta \psi_{\mathbf{i}_t}^\perp}) = 0\, ,\\
        (2)\quad &M_{\mathbf{i}_t,j}[\ket{\psi_{\mathbf{i}_t}} - 2 \lambda_j\ket{\partial_\eta \psi_{\mathbf{i}_t}^\perp}] = 0\, , \quad \mathrm{where} \quad \lambda_j = \frac{\braket{\psi_{\mathbf{i}_t}|M_{\mathbf{i}_t,j}^\dagger M_{\mathbf{i}_t,j} |\psi_{\mathbf{i}_t}}}{2\braket{\psi_{\mathbf{i}_t}|M_{\mathbf{i}_t,j}^\dagger M_{\mathbf{i}_t,j} |\partial_\eta \psi_{\mathbf{i}_t}^\perp}}\, ,
    \end{split}
\end{equation}
again for all measurement outcomes, and with $\ket{\partial_\eta \psi_{\mathbf{i}_t}^\perp} = \ket{\partial_\eta \psi_{\mathbf{i}_t}} - \ket{ \psi_{\mathbf{i}_t}}\braket{ \psi_{\mathbf{i}_t}|\partial_\eta \psi_{\mathbf{i}_t}}$. \\

We now show that conditions (1)-(2) [cf. Eq.~\eqref{eq:sat_cond_gen_pure}] are satisfied for the system-environment state with the corresponding measurements $\{\Pi_{\mathbf{i}_t,j}\}$ \textit{if} conditions (I) and (II) [cf. Eq.~\eqref{eq:conditions_saturation}] are fulfilled. Note that condition (I) corresponds to the orthogonality condition discussed in Ref.~\cite{godley_adaptive_2023}: $\braket{\Psi|\partial_\eta \Psi} = \tr(\rho (\partial_\eta H))= 0$, such that $\ket{\partial_\eta \Psi^\perp} = \ket{\partial_\eta \Psi}$. In contrast to Ref.~\cite{godley_adaptive_2023} where it was discussed that by adding a global (unphysical) and parameter dependent phase to the state $\ket{\Psi}$ one can recover the orthogonality, we construct our model from the start in a way such that this is fulfilled. This is due to the fact that the aim of our work was to show which type of models saturates the theoretical bound by typical classical measurements.
Further, imprinting such a global phase might be in general not possible. 
For condition (1) we therefore find 
$$
    \braket{\Psi|\Pi_{\mathbf{i}_t,j}^\dagger \Pi_{\mathbf{i}_t,j} |\partial_\eta \Psi^\perp} = \braket{\tilde{\psi}_{\mathbf{i}_t}|M_{\mathbf{i}_t,j}^\dagger M_{\mathbf{i}_t,j} |\partial_\eta \tilde{\psi}_{\mathbf{i}_t}}= \braket{\tilde{\psi}_{\mathbf{i}_t}|\tilde{\psi}_{\mathbf{i}_t}}\braket{\psi_{\mathbf{i}_t}|M_{\mathbf{i}_t,j}^\dagger M_{\mathbf{i}_t,j} |\phi_{\mathbf{i}_t}}\, ,
$$
where we used that the projector $\ket{\mathbf{i}_t}\bra{\mathbf{i}_t}$ selects a specific trajectory. With the definition of $\ket{\partial_\eta \psi_{\mathbf{i}_t}^\perp}$ and $\ket{\partial_\eta \psi_{\mathbf{i}_t}} = \ket{\phi_{\mathbf{i}_t}} - \mathrm{Re}(\braket{ \psi_{\mathbf{i}_t}|\phi_{\mathbf{i}_t}}) \ket{ \psi_{\mathbf{i}_t}}$ we find the relation 
$$
    \ket{\phi_{\mathbf{i}_t}} = \ket{\partial_\eta \psi_{\mathbf{i}_t}^\perp} +  \ket{ \psi_{\mathbf{i}_t}}\braket{\psi_{\mathbf{i}_t}| \phi_{\mathbf{i}_t}}\, ,
$$
such that
$$
    \braket{\Psi|\Pi_{\mathbf{i}_t,j}^\dagger \Pi_{\mathbf{i}_t,j} |\partial_\eta \Psi^\perp} = \braket{\tilde{\psi}_{\mathbf{i}_t}|\tilde{\psi}_{\mathbf{i}_t}}\left(\underbrace{\braket{\psi_{\mathbf{i}_t}|M_{\mathbf{i}_t,j}^\dagger M_{\mathbf{i}_t,j} |\partial_\eta \psi_{\mathbf{i}_t}^\perp}}_{\in \mathbb{R} \,\eqref{eq:sat_cond_gen_pure_sys}(1)}+\underbrace{\braket{\psi_{\mathbf{i}_t}|M_{\mathbf{i}_t,j}^\dagger M_{\mathbf{i}_t,j}  | \psi_{\mathbf{i}_t}}}_{p(j|\mathbf{i}_t)}\braket{\psi_{\mathbf{i}_t}| \phi_{\mathbf{i}_t}}\right)\, .
$$
Therefore, we find that 
$$
    \mathrm{Im}\left(\braket{\Psi|\Pi_{\mathbf{i}_t,j}^\dagger \Pi_{\mathbf{i}_t,j} |\partial_\eta \Psi^\perp}\right) = 0 \quad \Longleftrightarrow \quad \braket{\psi_{\mathbf{i}_t}| \phi_{\mathbf{i}_t}}\in \mathbb{R}\, .
$$
To the end of showing that condition (II) of Eq.~\eqref{eq:sat_cond_gen_pure} is fulfilled as well we use that
$$
    \lambda_{\mathbf{i}_t,j} = \frac{\braket{\Psi|\Pi_{\mathbf{i}_t,j}^\dagger \Pi_{\mathbf{i}_t,j} |\Psi}}{2\braket{\Psi|\Pi_{\mathbf{i}_t,j}^\dagger \Pi_{\mathbf{i}_t,j} |\partial_\eta \Psi^\perp}} = \frac{p(j|\mathbf{i}_t)}{2(\braket{\psi_{\mathbf{i}_t}|M_{\mathbf{i}_t,j}^\dagger M_{\mathbf{i}_t,j} |\partial_\eta \psi_{\mathbf{i}_t}^\perp}+p(j|\mathbf{i}_t)\braket{\psi_{\mathbf{i}_t}| \phi_{\mathbf{i}_t}})} \, .
$$
Thus, we find for 
$$
    \Pi_{\mathbf{i}_t,j}[\ket{\Psi} - 2 \lambda_{\mathbf{i}_t,j} \ket{\partial_\eta \Psi^\perp}] = M_{\mathbf{i}_t,j}\ket{\tilde{\psi}_{\mathbf{i}_t}} - \frac{p(j|\mathbf{i}_t)}{\braket{\psi_{\mathbf{i}_t}|M_{\mathbf{i}_t,j}^\dagger M_{\mathbf{i}_t,j} |\partial_\eta \psi_{\mathbf{i}_t}^\perp}+p(j|\mathbf{i}_t)\braket{\psi_{\mathbf{i}_t}| \phi_{\mathbf{i}_t}}}M_{\mathbf{i}_t,j}\ket{\tilde{\phi}_{\mathbf{i}_t}}
$$
to be zero, the condition
\begin{align*}
    0 &= \braket{\psi_{\mathbf{i}_t}|M_{\mathbf{i}_t,j}^\dagger M_{\mathbf{i}_t,j} |\partial_\eta \psi_{\mathbf{i}_t}^\perp} M_{\mathbf{i}_t,j}\ket{\psi_{\mathbf{i}_t}} + p(j|\mathbf{i}_t)\braket{\psi_{\mathbf{i}_t}| \phi_{\mathbf{i}_t}} M_{\mathbf{i}_t,j}\ket{\psi_{\mathbf{i}_t}} - p(j|\mathbf{i}_t)M_{\mathbf{i}_t,j}\ket{\phi_{\mathbf{i}_t}}\\\
    &= M_{\mathbf{i}_t,j}[\ket{\psi_{\mathbf{i}_t}} - 2 \lambda_j\ket{\partial_\eta \psi_{\mathbf{i}_t}^\perp}]\, ,
\end{align*}
which is just condition (2) for $M_{\mathbf{i}_t,j}$ [cf. Eq.~\eqref{eq:sat_cond_gen_pure_sys}].\\

\noindent Consequently, we have shown that the optimality conditions derived in Ref.~\cite{braunstein_statistical_1994} are fulfilled for the considered types of classical measurements, i.e. photon counting and homodyne detection, and models satisfying the conditions (I) and (II) given in Eq.~\eqref{eq:conditions_saturation}.\\ 
\end{document}